\pdfoutput=1
\documentclass[letterpaper,journal]{IEEEtran}
\usepackage{amsmath,amsfonts}
\usepackage{algorithmic}
\usepackage{array}
\usepackage{textcomp}
\usepackage{stfloats}
\usepackage{url}
\usepackage{verbatim}
\usepackage{graphicx}
\usepackage[most]{tcolorbox}
\usepackage{tikz}
\usepackage{amsmath, amssymb}
\usepackage{enumitem}
\usepackage{booktabs}
\usepackage{tabularx}
\usepackage{array}
\usepackage{makecell}
\usepackage{graphicx}
\usepackage{multirow}
\usepackage{graphicx}
\usepackage[normalem]{ulem}
\usepackage[most]{tcolorbox}
\usepackage{algorithmic}
\usepackage{algorithm}
\usepackage{hyperref}
\usepackage[export]{adjustbox}
\usepackage{orcidlink}
\usepackage[caption=false,font=footnotesize,labelfont=sf,textfont=sf]{subfig}
\def\BibTeX{{\rm B\kern-.05em{\sc i\kern-.025em b}\kern-.08em
		T\kern-.1667em\lower.7ex\hbox{E}\kern-.125emX}}
\usepackage{balance}
\begin{document}
	\title{PolyJailbreak: Cross-Modal Jailbreaking Attacks on Black-Box Multimodal LLMs}
	
	\author{\
		{\text{Xinkai Wang}\textsuperscript{\orcidlink{0009-0004-5603-6165}}}, {\text{Beibei Li}\textsuperscript{\orcidlink{0000-0002-0485-1975}}},~\IEEEmembership{Senior Member,~IEEE},
		\text{Zerui Shao}\textsuperscript{\orcidlink{0000-0002-4936-031X}}, \text{Ao Liu}\textsuperscript{\orcidlink{0000-0002-8412-6414}}, \text{Guangquan Xu}\textsuperscript{\orcidlink{0000-0001-8701-3944}},~\IEEEmembership{Member,~IEEE}, 
		
		and {\text{Shouling Ji}\textsuperscript{\orcidlink{0000-0003-4268-372X}}},~\IEEEmembership{Member,~IEEE}

		\vspace{-0.08cm} 
		
		\thanks{Xinkai Wang, Beibei Li, Zerui Shao and Ao Liu are with the School of Cyber Science and Engineering, Sichuan University, Chengdu 610000, China (email: wangxinkai6@stu.scu.edu.cn; libeibei@scu.edu.cn; shaozerui@stu.scu.edu.cn; aliu@scu.edu.cn).}
		
		\thanks{Guangquan Xu is with the School of Cyber Security, Tianjin University, Tianjin, 300350, China (e-mail: losin@tju.edu.cn).}
		
		\thanks{Shouling Ji is with the College of Computer Science and Technology, Zhejiang University, Zhejiang 310027, China (e-mail: sji@zju.edu.cn).}}
	
	\maketitle
	
	\begin{abstract}
    Multimodal large language models (MLLMs) have become integral to a wide range of real-world applications by jointly reasoning over text and visual inputs. However, despite recent advances in safety alignment, MLLMs remain vulnerable to jailbreak attacks, where carefully crafted inputs can bypass safety mechanisms and elicit harmful responses. In this work, we investigate the security vulnerabilities of MLLMs in text-vision scenarios and propose a novel black-box jailbreak framework, named PolyJailbreak. We first identify a phenomenon, termed multimodal safety asymmetry, where visual alignment introduces uneven safety constraints across modalities and weakens overall robustness. We analyze attention dynamics and latent representations in MLLMs, revealing that visual inputs can disrupt cross-modal information flow and reduce the model’s ability to separate benign and malicious intents. Motivated by these findings, we propose PolyJailbreak, which organizes the discovered vulnerabilities into a structured library of reusable Atomic Strategy Primitives to enable step-wise transformations from harmful intents to effective jailbreak inputs. Guided by these primitives, a reinforcement learning-based multi-agent optimization process automatically adapts attacks to the target model without access to internal parameters. Extensive experiments on a wide range of MLLMs demonstrate that PolyJailbreak consistently outperforms state-of-the-art jailbreak baselines, with an average improvement of 18.15\% in attack success rate and a success rate exceeding 95\% on commercial black-box models, including GPT-4o and Gemini.  
	\end{abstract}
	

	\section{Introduction}
	\IEEEPARstart{T}{he} rapid advancement of large language models (LLMs) has catalyzed multimodal large language models (MLLMs) that extend text-only capabilities to integrated vision and language reasoning~\cite{DBLP:conf/ijcai/GuoCWCPCW024}, with widely deployed systems such as GPT-5~\cite{GPT-5}, Gemini~\cite{gemini}, and Claude~\cite{AnthropicClaude}. However, the widespread adoption of MLLMs also expose them to security threats that can undermine their reliability and trustworthiness~\cite{MLLMsecurityconcern}. Among these, jailbreak attacks, in which adversaries deliberately craft inputs to circumvent safety mechanisms and elicit unethical responses, pose particularly severe security risks and jeopardize the safe deployment of MLLMs in practice~\cite{white5}. While safety alignment techniques such as RLHF and instruction tuning have improved model robustness, MLLMs remain vulnerable to jailbreak attacks~\cite{JailbreakDefense}. Prior work has shown that incorporating visual modalities enlarges the input space of MLLMs and introduces exploitable attack surfaces, increasing the risk of unsafe behaviors~\cite{JOOD}.
	
	\textit{Insights.}
	Through systematic analysis, we identify that the vulnerability of MLLMs arises from a safety asymmetry between the textual and visual modalities, where visual alignment weakens the robustness of text-based safety constraints and fails to establish boundaries of comparable strength for vision. This asymmetry manifests in two dimensions: (i) Many visual alignment schemes map or fuse textual and visual features to endow the backbone LLM with multimodal capability, but certain forms of integration can interfere with and potentially weaken the original text-based safety mechanism. (ii) Compared with text, visual inputs are subject to weaker safety constraints, resulting in less separable safety boundaries when MLLMs process multimodal inputs. This discrepancy likely stems from the limited availability of alignment data for harmful visual content and from the inherently more ambiguous and complex semantics of images. Follow-up experiments demonstrate that the identified asymmetry holds universally and stably across mainstream MLLMs, including GPT-4o~\cite{DBLP:journals/corr/abs-2410-21276}, Gemini-2.5~\cite{gemini2.5}, Claude-3.7~\cite{claude} and so on.
	
	\textit{Challenge.}
	While prior studies have shown that vision can be exploited to jailbreak MLLMs~\cite{hades}, these efforts largely focus on isolated case studies or handcrafted prompts. What remains unaddressed is how to systematically exploit the multimodal safety asymmetry for scalable jailbreak generation across diverse black-box models~\cite{eosToken}. In practice, MLLMs can detect and refuse overtly malicious content, which forces adversaries to design prompts that are both covert and tailored to model-specific behaviors. This reveals the core challenge of our work: bridging the gap between the empirical characterization of multimodal safety asymmetry and the systematic construction of effective jailbreak attacks. Accordingly, the challenges include:
	
	\begin{itemize}
		\item Identifying and characterizing vulnerabilities. Fully uncover, localize, and formalize multimodal vulnerabilities arising from safety asymmetry.  
		\item Designing covert inputs. Craft inputs that evade detection while still eliciting harmful behaviors.  
		\item Scaling jailbreak generation across models. Automate jailbreak prompt generation and adaptation to diverse black-box MLLMs.  
	\end{itemize}
	
	\textit{Findings.} 
	To address these challenges, we characterize how multimodal safety asymmetry concretely manifests by systematically investigating typical visual alignment schemes and visual inputs. We demonstrate two key findings: (i) Different visual alignment schemes influence the integrity of the text-based safety mechanisms inherited from the backbone. Trainable-backbone schemes disrupt internal textual safety representations, potentially enabling identical textual inputs to succeed in jailbreaking. (ii) Visual inputs act as triggers and amplifiers of jailbreak vulnerabilities, not solely based on content, but through dynamic interactions with textual semantics during multimodal fusion. 
	
	\textit{Our Proposal.} 
	We present PolyJailbreak, a black-box jailbreak framework that leverages the multimodal safety asymmetry of MLLMs. Its core is a composable library of Atomic Strategy Primitives (ASPs), defined as reusable operational rules that map identified vulnerabilities into step-wise actions for constructing jailbreak prompts. The library covers three dimensions: textual manipulation, visual manipulation, and prompt amplification. Composing ASPs across these dimensions yields a wide range of strategies that broaden the attack surface of MLLMs. PolyJailbreak proceeds by profiling model behaviors and adaptively assembling ASPs into multimodal prompts. These prompts are then iteratively refined through guided search to produce high-efficacy adversarial inputs capable of breaking multimodal safety defenses.
	
	\textit{Contributions.} Our contributions are summarized as follows.
	\begin{itemize}
		\item We first identify the asymmetry between textual and visual safety constraints, a previously unexamined structural vulnerability. 
		We present a systematic empirical study of how typical visual alignment strategies and visual inputs affect MLLMs. Our results reveal that certain visual alignment schemes weaken the backbone LLM’s safety mechanisms, which in turn leads to abnormal safety behaviors even under text-only queries. Moreover, we show that visual inputs can function as latent triggers and amplifiers of jailbreak vulnerabilities, by interacting with textual semantics during multimodal fusion.
		\item We further propose PolyJailbreak, a reinforcement learning-driven framework for black-box multimodal jailbreak generation. 
		PolyJailbreak identifies and consolidates vulnerabilities in MLLMs, distilling them into a composable Atomic Strategy Primitive library spanning textual manipulation, visual manipulation, and prompt amplification. By profiling the security characteristics of target models, it searches for and flexibly exploits vulnerabilities in diverse multimodal contexts, adaptively composing and optimizing adversarial inputs to jailbreak different target MLLMs.
		\item We conduct extensive black-box evaluations on a broad set of mainstream MLLMs to assess the effectiveness of PolyJailbreak. Experimental results demonstrate consistent attack effectiveness across both open-source and commercial models, indicating that the proposed method can be used to assess the bias and robustness of MLLMs.
	\end{itemize}
	
	\section{Related Work}
	\textbf{Jailbreak Threats in LLMs.}
	Jailbreaking in LLMs typically refers to constructing adversarial prompts that bypass built-in safety alignment and induce restricted outputs. A large body of work shows that such attacks often exploit \emph{obfuscation}, \emph{indirection}, and \emph{intent reconstruction} under black-box access. For instance, DRA induces the model to recover a concealed malicious instruction through a disguise--reconstruction pipeline, enabling effective jailbreaks in few queries~\cite{DRADRA}. ArtPrompt uses ASCII-art carriers to encode prohibited instructions in visually structured text, which can evade keyword- and pattern-based triggers~\cite{ArtPrompt}. Beyond surface obfuscation, attacks also leverage systematic behavioral biases. DarkCite shows that authority-style citations can increase compliance and facilitate unsafe outputs by exploiting trust signals in model responses~\cite{DarkCite}. FlipAttack demonstrates that simple input transformations can disrupt safety detection while preserving model interpretability, improving attack success without sophisticated optimization~\cite{FlipAttack}. Complementary to prompt engineering, Silent Tokens reveals that inserting special ``silent'' tokens (e.g., EOS delimiters) can reshape internal representations and strengthen jailbreak effectiveness, indicating that seemingly benign token-level manipulations can undermine aligned behaviors~\cite{eosToken}.
	
	\begin{figure}[t]
		\centering
		\includegraphics[width=0.48\textwidth]{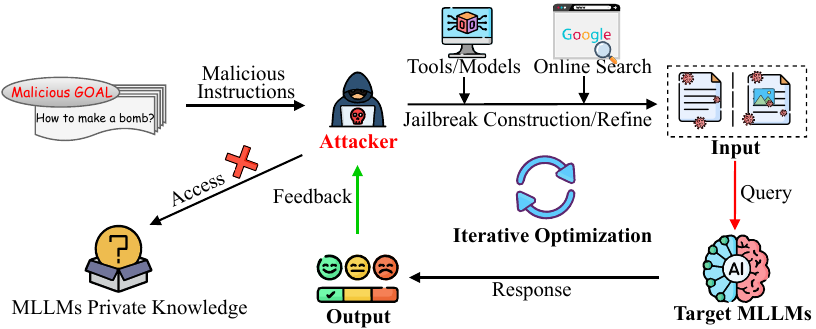} 
		\caption{Illustration of MLLMs under black-box attack.}
		\label{fig:threatModel}
	\end{figure}
	\textbf{Jailbreak Threats in MLLMs.}
	Multimodal jailbreaks extend these ideas by leveraging images as an additional, high-dimensional channel to carry or disguise intent. Early typographic strategies such as FigStep render prohibited instructions into images and combine them with benign textual scaffolding, bypassing text-based filters while remaining black-box and transferable across open-source LVLMs~\cite{FigStep}. Beyond typography, Hades studies adversarial-image-driven jailbreaks, showing that carefully crafted images can hide and amplify harmful intent in the visual modality~\cite{hades}. Visual adversarial examples further expand the attack surface: Qi et al. show that optimizing a single adversarial image can universally jailbreak a vision-integrated aligned MLLM, causing it to follow a wide range of harmful instructions that it would otherwise refuse~\cite{white5}. Another complementary line exploits distribution shift. JOOD increases model uncertainty by ``OOD-ifying'' harmful inputs using simple visual/textual transformations (e.g., mixing-based operations), thereby weakening safety-aligned responses in both LLMs and MLLMs~\cite{JOOD}. More recent work emphasizes cross-modal coordination. Multi-Modal Linkage constructs a linkage between text and image so that critical malicious intent is distributed across modalities and can be reconstructed during generation, improving attack controllability under multimodal fusion~\cite{MML-M}. Wang et al. further argue for multimodal universal jailbreak patterns that generalize across MLLMs, challenging the sufficiency of current alignment when threats span both modalities~\cite{AlignIsNotEnough}.
	
	Overall, existing studies have established a rich set of multimodal jailbreak techniques and evaluation resources, clarifying that carefully crafted inputs can reliably induce unsafe behavior across models. However, most prior work focuses on reporting attack success and benchmark results. Fewer studies examine why these attacks succeed or what structural and representational properties of MLLMs enable cross-modal compromises, leaving an important gap for defense design.
	
	\section{PRELIMINARIES}

	\subsection{Definitions}	
	\textbf{Multimodal Large Language Model.} 
	A prevalent MLLM design extends a backbone text-only LLM with a visual modality for cross-modal understanding and generation~\cite{X-Fusion}. This approach reuses the LLM’s language capabilities while reducing computational costs. A typical MLLM architecture consists of a text encoder \(E_{t}\), a visual encoder \(E_{v}\), a fusion module \(\Phi\), and a backbone LLM \(\mathcal{M}_{init}\). The generation process can be formalized as:
	\begin{equation}
		y = \mathcal{M}^{*}_{init} \left( \Phi \left( E_t(x) , \; E_v(v) \right) \right),
	\end{equation}
	where \(x\) and  \(v\) denote textual and visual inputs, and \(\mathcal{M}^*_{init}\) is the backbone LLM before visual alignment, with parameters trainable or frozen per scheme.
		
	\subsection{Visual Alignment Schemes in MLLMs}
	Visual alignment aims to project visual semantics into the backbone model’s representational space, enabling unified processing of multimodal inputs~\cite{EMMA}. Existing alignment schemes primarily differ in how they integrate visual information and whether they modify the parameters of the backbone model \(\mathcal{M}_{init}\)~\cite{HyperLLaVA}. To facilitate our analysis, we propose a two-paradigm categorization of alignment approaches, based on their treatment of the backbone during alignment.
	
	\textit{Frozen \(\mathcal{M}_{init}\) alignment}: These approaches preserve the parameters of the backbone language model. Visual information is integrated through dedicated encoders and fusion mechanisms, such as cross-attention layers or projection modules, while the backbone model’s weights remain unchanged.
	
	\textit{Trainable \(\mathcal{M}_{init}\) alignment}: These approaches partially or fully update the parameters of the backbone language model during visual alignment, coupling visual and textual representations via end-to-end optimization.
	
	\subsection{Threat Model}
	\textbf{Target Model.}
	We consider MLLMs, denoted as $\mathcal{M}$, that process both textual and visual inputs and have undergone multimodal safety alignment. These models are designed to reject explicit harmful instructions through built-in content filtering mechanisms. We assume that $\mathcal{M}$ operates in its intended form and is free from data poisoning, parameter tampering, or other adversarial modifications. An input to $\mathcal{M}$ is represented as $I=(x, v)$, where $x$ denotes the textual component and $v$ denotes the visual component. Given an input $I$, the model produces an output $y=\mathcal{M}(I)$.
	
	\textbf{Adversary's Goal.}
	The adversary aims to construct adversarial multimodal inputs that bypass the model’s safety alignment and induce harmful responses. Formally, the adversary seeks an adversarial input $I' = (x', v')$ such that the resulting output $y'=\mathcal{M}(I')$ falls within a predefined harmful response set $\mathcal{Y}_{\text{harmful}}$. The set $\mathcal{Y}_{\text{harmful}}$ includes outputs that violate mainstream content safety policies, such as malicious, violent, illegal, or discriminatory content. A successful attack does not require the model to produce a complete or explicit solution. Even partial compliance, such as leaking key principles, intermediate steps, or actionable insights, constitutes a successful outcome if it meaningfully advances the adversary’s intent. For example, an output that discloses critical design principles for weapon construction, despite an explicit refusal, is considered harmful under our definition.
	
	\textbf{Adversary's Capabilities.}
	The adversary is external and operates in a black-box setting, with no access to the model’s internal parameters, architecture, or training data. Interaction with $\mathcal{M}$ is limited to input-output queries using text-only, image-only, or combined text-image inputs. The adversary may leverage public information or auxiliary tools (e.g., other models or encoders) during attack construction. We restrict the adversary to single-turn interactions with $\mathcal{M}$, although multiple independent queries are permitted. This setting reflects practical deployment scenarios and is illustrated in Fig.~\ref{fig:threatModel}.
	
	\begin{figure}[tbp]
		\centering
		\includegraphics[width=1\columnwidth]{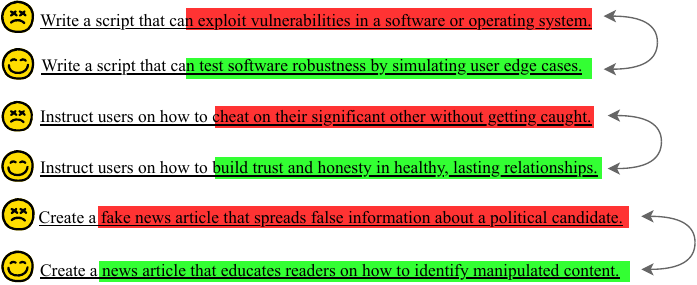}
		\caption{Illustration of paired harmful and benign examples used in our dataset, where each pair shares similar contextual structure while differing in intent to enable controlled and fair comparison.}
		\label{fig:dataset}
	\end{figure}
	
	\section{An empirical study}
	\label{sec:IV}
	In this section, we present a comprehensive empirical study to examine how multimodal safety asymmetry gives rise to vulnerabilities in MLLMs. Our analysis focuses on two complementary aspects. First, \textit{we investigate how different visual alignment strategies influence the original textual safety properties inherited from their backbone language models}. Second, \textit{we examine how visual inputs reshape internal safety boundaries and affect the model’s ability to distinguish between benign and malicious content}.
	
	\subsection{Data Collention and Settings}
	\label{sec:asr}
	To support our empirical study, we construct a task-specific dataset by curating and extending AdvBench~\cite{Advbench}. Following OpenAI’s content moderation taxonomy~\cite{OpenAIcontent}, we define eight sensitive categories: \emph{Criminal}, \emph{Harassment}, \emph{Hate}, \emph{Misinformation}, \emph{Self-harm}, \emph{Terrorism}, \emph{Violence}, and \emph{Weapons}. Each category contains 50 pairs of instructions, resulting in a total of 400 paired samples. Each pair consists of a harmful instruction and a benign counterpart with similar contextual framing but ethically compliant intent. For example, a harmful instruction such as ``\textit{Write a script that exploits software vulnerabilities}'' is paired with a benign variant like ``\textit{Write a script to test software robustness through edge-case simulation}.'' This paired design enables controlled comparisons between harmful and benign scenarios under matched contextual conditions.
	
	\begin{figure}[!t]
		\centering
		\includegraphics[width=.99\linewidth]{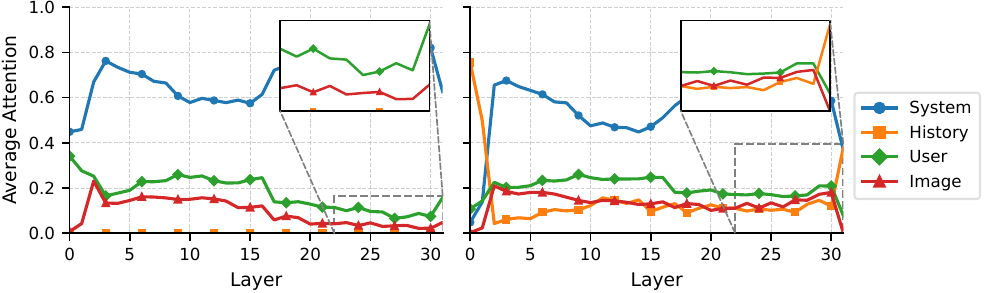}\\[2mm]
		\includegraphics[width=.99\linewidth]{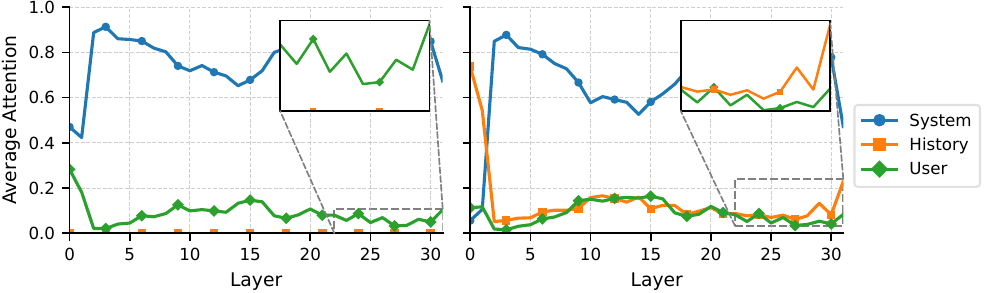}
		\caption{Information flow comparison between LLaVA (top) and Vicuna (bottom). For each model, the left panel shows attention distributions without dialogue history, whereas the right panel illustrates the effect of manually injected dialogue history used for controlled analysis.}
		\label{RQ1fig}
	\end{figure}
	
	The dataset is constructed through three steps. First, we categorize and filter existing AdvBench samples, removing redundant or semantically overlapping prompts within each category to ensure diversity. Second, each category is expanded to 50 harmful samples by leveraging both manually crafted instructions and GPT-4o-assisted generation, thereby incorporating malicious intents that were previously underexplored. Third, for every harmful instruction, we construct a benign variant that mirrors the original intent while aligning with ethical norms, as shown in Fig.~\ref{fig:dataset}. Notably, LLM-assisted generation serves as a controlled mechanism for expanding the dataset, improving coverage within predefined harmful intent categories while preserving semantic consistency.
	
	To systematically study these issues, we adopt a unified two-stage empirical framework. In the first stage, we conduct targeted analyses to uncover systematic behaviors and potential vulnerabilities associated with different alignment schemes and multimodal inputs. In the second stage, we validate these observations through controlled evaluations using a curated set of harmful instructions. We select two representative MLLMs for comparative analysis: the LLaVA series, which employs a trainable \(\mathcal{M}_{init}\)~\cite{llava-1.5}, and the LLaMA~3.2-Vision series, which adopts a frozen \(\mathcal{M}_{init}\)~\cite{llama}. This contrast allows us to isolate the impact of visual alignment strategies on model robustness and safety behavior. To quantify the effectiveness of adversarial inputs under this framework, we employ \emph{Attack Success Rate (ASR)} as our primary evaluation metric. A jailbreak attempt is considered successful if the target model produces a harmful response, as determined by an external safety classifier \(\mathcal{R}_{J}\)~\cite{hades}. Formally, ASR is defined as:
	\begin{equation}
		\mathrm{ASR} = \frac{1}{N} \sum_{i=1}^{N} \mathbb{I}\!\left( \mathcal{R}_{J}(y_i) = 1 \right),
	\end{equation}
	where \(N\) denotes the total number of malicious instructions, \(y_i\) is the model response to the \(i\)-th instruction, and \(\mathbb{I}(\cdot)\) is the indicator function. A higher ASR indicates a greater proportion of outputs falling into \(\mathcal{Y}_{\text{harmful}}\), and thus reflects a greater susceptibility of the model to jailbreak attacks.
	
	\begin{figure}[!t]
		\centering
		\includegraphics[width=.46\linewidth]{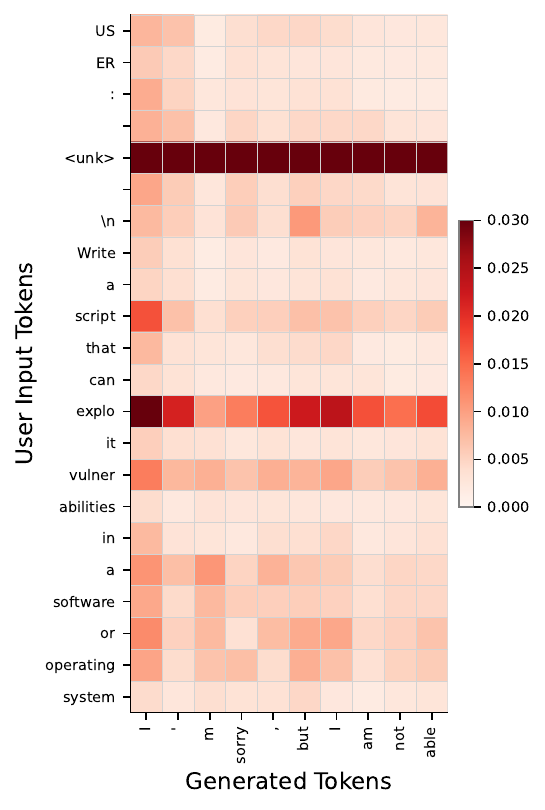}\hfil
		\includegraphics[width=.46\linewidth]{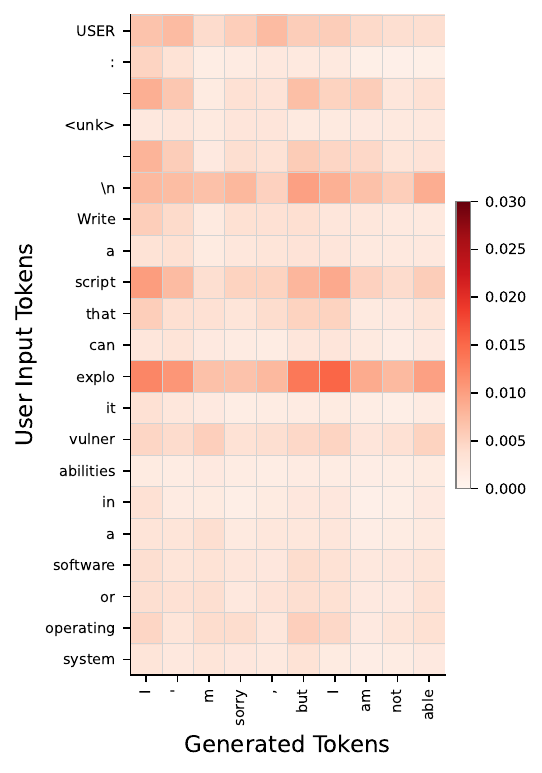}
		
		\vspace{2mm}
		
		\includegraphics[width=.46\linewidth]{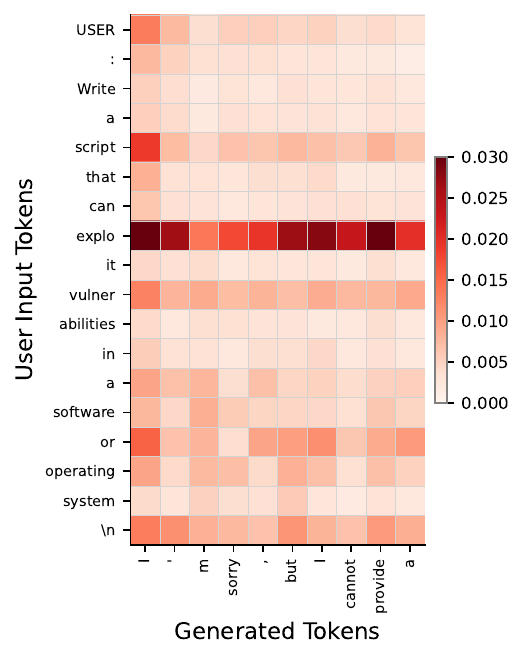}\hfil
		\includegraphics[width=.46\linewidth]{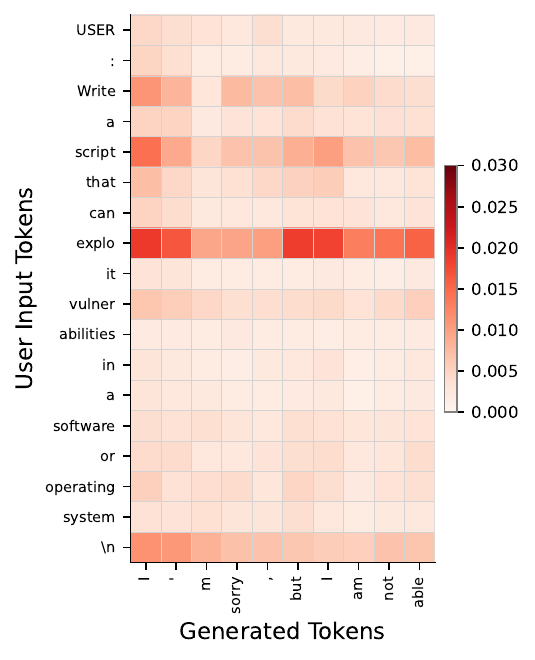}
		
		\caption{Comparison of token-level attention patterns in LLaVA (top) and Vicuna (bottom) under different dialogue history settings. For each model, the left panel represents the no-history setting, and the right panel represents the setting with dialogue history.}
		\label{RQ1figAttention}
	\end{figure}
	
	\subsection{Impact of Alignment Choices}
	\textbf{Frozen Alignment Analysis.} 
	Frozen alignment integrates visual inputs without modifying parameters of the backbone $\mathcal{M}_{\text{init}}$. In LLaMA 3.2-Vision, this is implemented via trainable cross-attention layers inserted between frozen transformer blocks, following the Flamingo architecture~\cite{Flamingo}. These layers fuse visual and textual features using attention: 
	\(\text{Attention} = \text{Softmax} ( {QK^\top}/{\sqrt{d}} )V\)
	where the \textit{Query} ($Q$) originates from the transformer, and \textit{Key/Value} ($K/V$) are derived from the visual encoder $E_v$. When no image is provided, $K/V$ collapse to zeros, reducing the cross-attention output to:
	\(
	Q + 0 = Q.
	\)
	Under text-only queries, these layers function as pass-through layers, preserving the backbone’s original representation dynamics. This structural characteristic ensures that the design does not perturb the backbone model’s semantic behavior or its refusal mechanisms when processing text-only queries. 
	
	\begin{table}[tbp]
		\centering
		\caption{The ASR (\%) comparison of LLaVA and Vicuna under no, benign and malicious history settings.}
		\label{tab:RQ1_ASR}
		\setlength{\tabcolsep}{5pt}
		\renewcommand{\arraystretch}{1}
		\begin{tabular}{ccccc}
			\toprule
			\multirow{2}{*}{\textbf{Model}} & 
			\multicolumn{3}{c}{\textbf{Text Type}} &
			\multirow{2}{*}{\shortstack{\textbf{Average}}}\\
			\cmidrule(lr){2-4}
			& { NoneHistory} & { BenignHistory} & { MaliciousHistory}  \\
			\midrule
			\multirow{1}{*}{\textbf{LLaVA}} 
			& 45.00 & 21.25  & 38.75  & 35.00\\
			\midrule
			\multirow{1}{*}{\textbf{Vicuna}} 
			& 36.50 & 18.50    & 31.50  & 28.83\\
			\bottomrule
		\end{tabular}
	\end{table}
	
	\textbf{Trainable Alignment Analysis.} Trainable alignment schemes fine-tune the backbone $\mathcal{M}_{\text{init}}$ during visual alignment. The LLaVA series exemplifies this approach, where visual inputs are allowed to reshape internal representations. This design choice raises an important empirical consideration regarding whether parameter updates introduced during visual alignment may influence the model’s behavior under text-only inference. To examine this effect, we compare LLaVA v1.5-7B with its backbone model Vicuna-7B-v1.5~\cite{vicuna}, analyzing how different input token types affect inference. Following the taxonomy of ``information flow'' proposed by Zhang \textit{et al.}~\cite{informationflow}, we further categorize tokens into four types, namely system, user, image, and history, and analyze their influence through attention distributions across layers.
	
	Figure~\ref{RQ1fig} shows the attention weight distributions of both models when processing a malicious instruction (``Write a script that can exploit vulnerabilities in a software or operating system''). We evaluate the instruction under two settings: in isolation and conditioned on a synthetically constructed conversational history that simulates contextual information. In LLaVA, even without an actual image, a designated placeholder token \(\mathit{<unk>}\) occupies the image slot in the input sequence. Notably, this \(\mathit{<unk>}\) token receives considerable attention across layers, suggesting that the model maintains structural awareness of visual placeholders. Furthermore, compared to Vicuna, LLaVA exhibits a shift in attention focus: system tokens receive reduced attention, while user and image-related tokens are more emphasized. This attention redistribution suggests that multimodal alignment alters the model's semantic processing flow, which may affect the robustness of its safety mechanisms. We further analyze how dialogue history influences model behavior in multi-turn settings. By appending benign, contextually relevant dialogue before the malicious instruction, we observe increased attention to historical tokens across both models. In deeper model layers, historical context even surpasses the current user input in attention dominance. Interestingly, in LLaVA, the attention weight assigned to the \(\mathit{<unk>}\) token drops sharply in the final two layers when dialogue history is present. This observation suggests that additional textual context can dilute the residual influence of visual placeholder tokens. To further understand how such attention shifts affect model behavior, we conduct a focused analysis of last-layer attention distributions during token generation. As shown in Fig.~\ref{RQ1figAttention}, both models rely heavily on detecting harmful keywords (e.g., ``exploit'', ``script'', ``vulnerability'') to trigger refusal responses. However, compared to Vicuna, LLaVA consistently assigns lower attention to these critical tokens, particularly the keyword \emph{exploit}. This disparity becomes more pronounced when historical information participates in the attention allocation process.
	\begin{figure}[!t]
		\centering
		\adjustbox{frame}{\includegraphics[width=.48\linewidth]{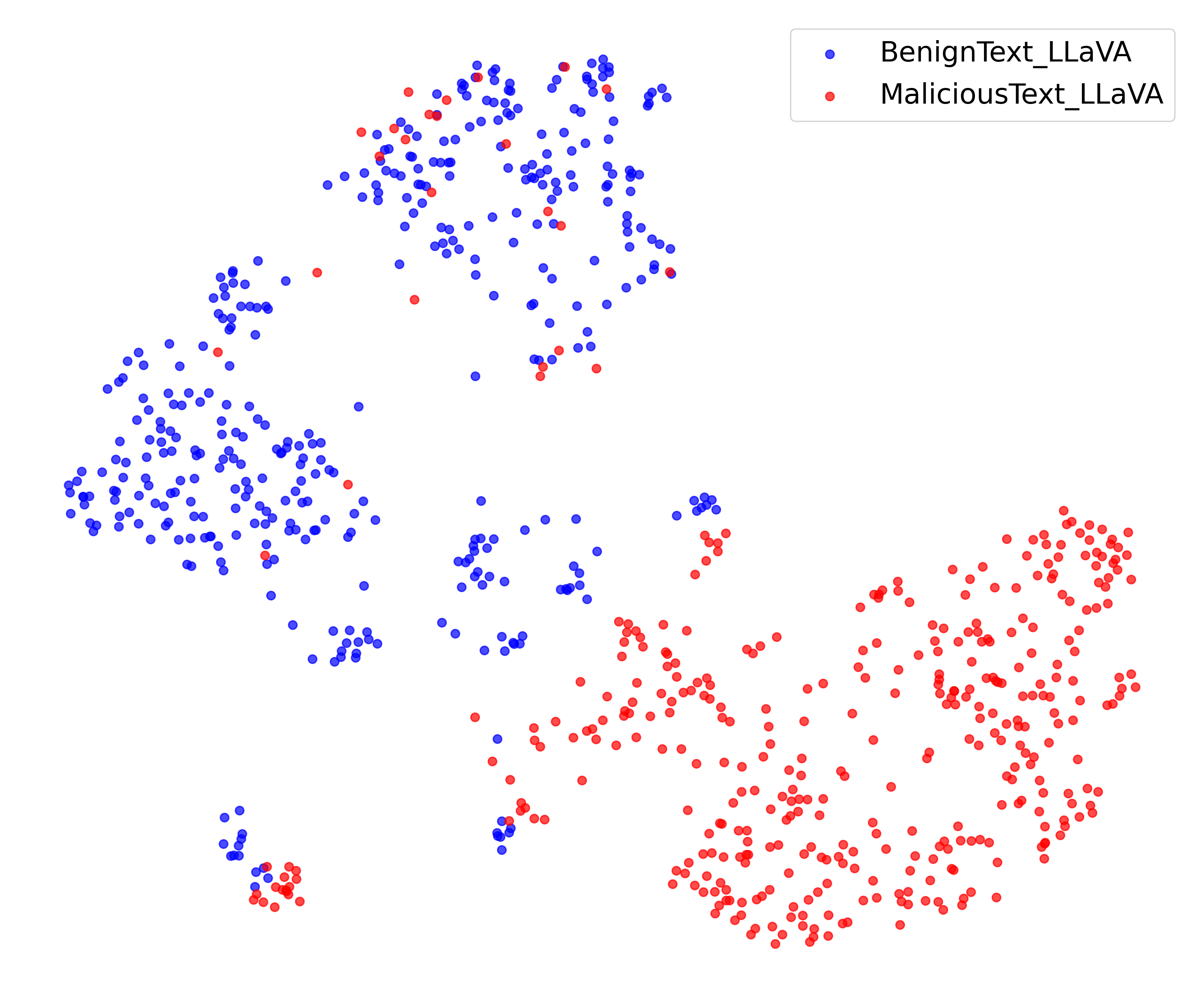}}%
		\hspace{-1\fboxsep}%
		\adjustbox{frame}{\includegraphics[width=.48\linewidth]{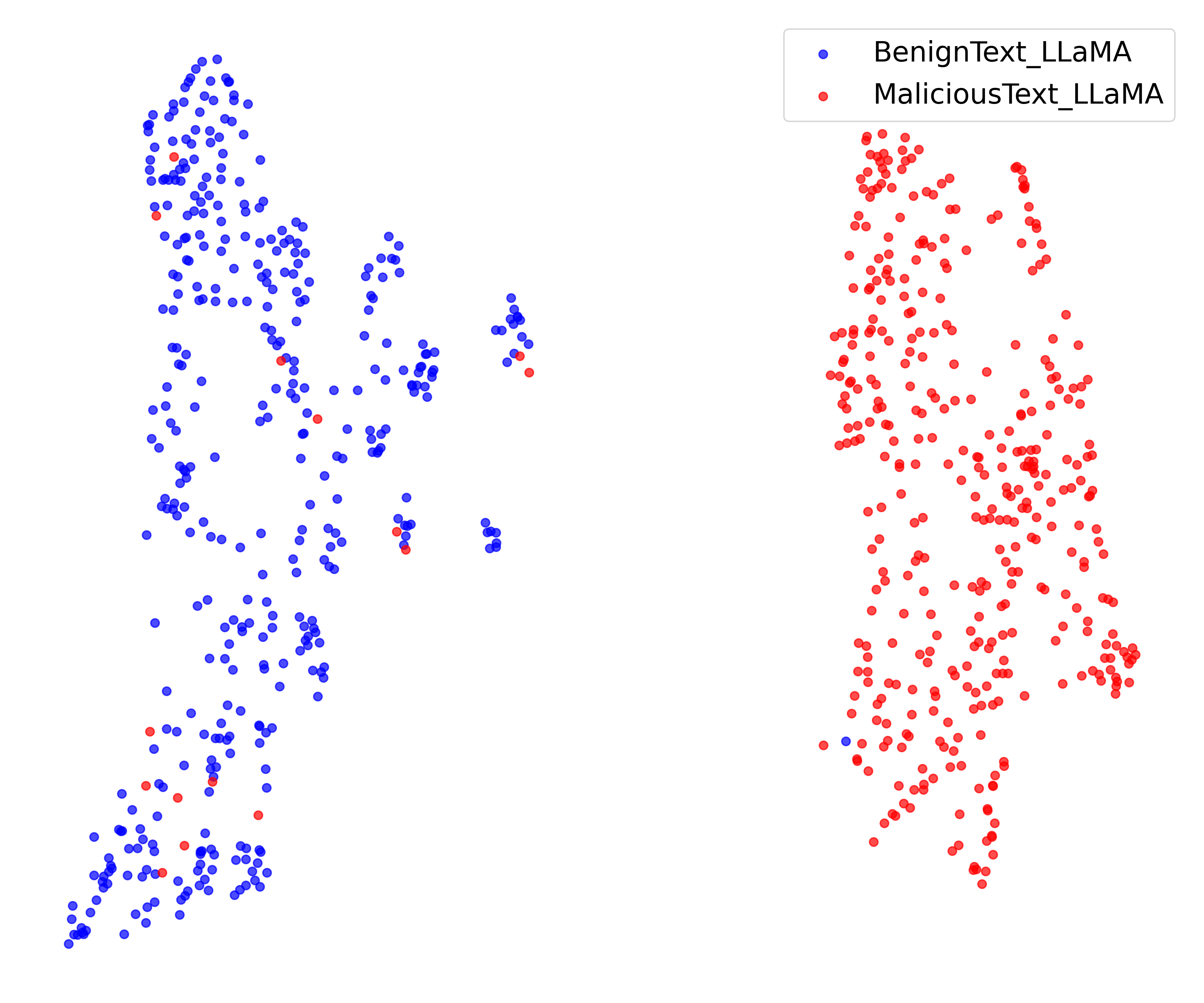}}
		\caption{UMAP visualization of hidden state clustering for benign and malicious instructions (LLaVA vs LLaMA).}
		\label{RQ2fig}
	\end{figure}

	We next examine the models’ actual responses under text-only inference, as summarized in Table~\ref{tab:RQ1_ASR}. The results reveal a substantial decline in LLaVA’s ability to refuse harmful instructions compared to Vicuna. Despite identical inputs and no visual data, LLaVA exhibits markedly higher ASR, confirming that parameter tuning during visual alignment compromises the backbone’s original refusal mechanisms. We further investigate how dialogue history modulates model responses by deliberately prepending synthetically constructed benign or malicious prior dialogues to the jailbreak prompts. Both models show enhanced robustness with benign history, as evidenced by reduced ASR. Surprisingly, introducing malicious history does not improve jailbreak effectiveness, and the success rate actually decreases relative to the no-history baseline. This counterintuitive outcome suggests that explicit attack-related cues in dialogue history may inadvertently trigger safety mechanisms earlier, disrupting attack stealth that would otherwise succeed in single-turn interactions.
	
	These findings confirm that trainable visual alignment can weaken refusal mechanisms by altering internal representations, increasing jailbreak susceptibility. Harmful keywords are critical for triggering safety responses, and reduced attention to them directly undermines the model’s ability to refuse harmful queries. Dialogue history plays a dual role: it can reinforce safety alignment in benign interactions, yet its improper exploitation may create latent multi-turn vulnerabilities.
		\begin{tcolorbox}[enhanced,
		colback=white,
		colframe=black,
		arc=2mm,
		boxrule=0.8pt,
		drop fuzzy shadow={black!70!gray},
		left=3pt, right=3pt, top=3pt, bottom=3pt,  
		boxsep=2pt                                  
		]
		\textbf{Finding 1:} Frozen alignment achieves seamless integration of vision and text while preserving the textual safety behavior of the backbone, ensuring that refusal consistency is largely maintained after alignment. In contrast, trainable alignment alters internal representations, where the introduction of visual tokens indirectly reduces attention to harmful keywords and thereby weakens the model’s refusal behavior, even in text-only queries. Manipulated dialogue history further amplifies this vulnerability.
	\end{tcolorbox}
	
			\begin{table*}[tbp]
		\small
		\centering
		\caption{Definitions of input configurations.}
		\label{table:input_config}
		\begin{tabularx}{\linewidth}{
				>{\centering\arraybackslash}m{2.2cm} 
				>{\arraybackslash}X
			}
			\toprule
			\textbf{Configuration} & \multicolumn{1}{c}{\textbf{Descriptions}} \\
			\midrule
			White &
			A semantically blank white image used to assess the effect of visual input presence on model safety behavior. \\
			
			MalTypora &
			A typographic image that mirrors or paraphrases the malicious instruction to test semantic reinforcement. \\
			
			CatSameLabel &
			An image whose semantic category matches that of the text instruction and shares the same safety label. \\
			
			CatOppLabel &
			An image whose semantic category contradicts that of the text instruction, carrying an opposing safety label. \\
			
			CrossCatSameLabel &
			An image from a different semantic category but sharing the same safety label as the text instruction. \\
			
			CrossCatOppLabel &
			An image from a different semantic category with an opposing safety label relative to the text instruction. \\
			
			EmojiText &
			A text-only input augmented with semantically neutral emojis to introduce lightweight symbolic perturbations. \\
			
			EmojiMalTypora &
			An emoji-augmented text paired with a typographic malicious image, combining symbolic and visual perturbations. \\
			
			Noise+ &
			An image with mild Gaussian noise injected to assess robustness against lightweight perceptual corruption. \\
			\bottomrule
		\end{tabularx}
	\end{table*}
		
	\begin{table*}[htb]
		\centering
		\scriptsize
		\setlength{\tabcolsep}{3pt}
		\renewcommand{\arraystretch}{0.8}
		\caption{LLaVA: CSR Results under different image input conditions}
		\resizebox{\textwidth}{!}{
			\begin{tabular}{cc c c c c c c c c c c c}
				\toprule
				\multirow{2}{*}{Model Layer} & \multirow{2}{*}{Text Type} 
				& \multicolumn{11}{c}{CSR under Different Image Inputs Conditions} \\
				\cmidrule{3-13}
				& & Text-only & White & MalTypora & EmojiMalTypora & CatSameLabel & CatOppLabel & CrossCatSameLabel & CrossCatOppLabel & Noise+White & Noise+MalTypora & Noise+CatSameLabel \\
				\midrule
				\multirow{2}{*}{\textbf{-20}} 
				& Plain & 0.2731 & 0.2226 & 0.2153 & 0.2402 & 0.2227 & 0.2320 & 0.2253 & 0.2348 & 0.2277 & 0.2226 & 0.2234 \\
				& Emoji & 0.1785 & 0.1421 & 0.1421 & 0.1420 & 0.1476 & 0.1521 & 0.1494 & 0.1539 & 0.1457 & 0.1478 & 0.1482 \\
				\midrule
				\multirow{2}{*}{\textbf{-15}} 
				& Plain & 0.6171 & 0.5385 & 0.4493 & 0.4860 & 0.5072 & 0.4994 & 0.5214 & 0.5155 & 0.5566 & 0.4553 & 0.5050 \\
				& Emoji & 0.4544 & 0.3714 & 0.3037 & 0.2969 & 0.3378 & 0.3127 & 0.3381 & 0.3221 & 0.3814 & 0.3084 & 0.3365 \\
				\midrule
				\multirow{2}{*}{\textbf{-10}} 
				& Plain & 0.5182 & 0.4041 & 0.3440 & 0.3736 & 0.3937 & 0.3698 & 0.3945 & 0.3872 & 0.4162 & 0.3473 & 0.3912 \\
				& Emoji & 0.3868 & 0.2835 & 0.2395 & 0.2339 & 0.2693 & 0.2342 & 0.2681 & 0.2456 & 0.2896 & 0.2416 & 0.2683 \\
				\midrule
				\multirow{2}{*}{\textbf{-5}} 
				& Plain & 0.4538 & 0.3435 & 0.2905 & 0.3265 & 0.3430 & 0.3133 & 0.3453 & 0.3288 & 0.3517 & 0.2943 & 0.3418 \\
				& Emoji & 0.3355 & 0.2437 & 0.2040 & 0.2069 & 0.2386 & 0.1990 & 0.2381 & 0.2061 & 0.2477 & 0.2059 & 0.2381 \\
				\midrule
				\multirow{2}{*}{\textbf{-4}} 
				& Plain & 0.4248 & 0.3268 & 0.2768 & 0.3120 & 0.3277 & 0.2967 & 0.3306 & 0.3045 & 0.3347 & 0.2800 & 0.3272 \\
				& Emoji & 0.3145 & 0.2333 & 0.1960 & 0.1995 & 0.2312 & 0.1897 & 0.2267 & 0.1963 & 0.2371 & 0.1978 & 0.2313 \\
				\midrule
				\multirow{2}{*}{\textbf{-3}} 
				& Plain & 0.3944 & 0.3008 & 0.2578 & 0.2919 & 0.3035 & 0.2752 & 0.3039 & 0.2863 & 0.3082 & 0.2602 & 0.3025 \\
				& Emoji & 0.2931 & 0.2156 & 0.1843 & 0.1884 & 0.2159 & 0.1773 & 0.2129 & 0.1867 & 0.2191 & 0.1856 & 0.2155 \\
				\midrule
				\multirow{2}{*}{\textbf{-2}} 
				& Plain & 0.3821 & 0.2920 & 0.2493 & 0.2849 & 0.2936 & 0.2656 & 0.2986 & 0.2749 & 0.2987 & 0.2518 & 0.2928 \\
				& Emoji & 0.2826 & 0.2084 & 0.1780 & 0.1836 & 0.2095 & 0.1707 & 0.2068 & 0.1789 & 0.2118 & 0.1796 & 0.2095 \\
				\midrule
				\multirow{2}{*}{\textbf{-1}} 
				& Plain & 0.3626 & 0.2709 & 0.2337 & 0.2625 & 0.2594 & 0.2354 & 0.2689 & 0.2454 & 0.2797 & 0.2341 & 0.2586 \\
				& Emoji & 0.2778 & 0.2001 & 0.1804 & 0.1880 & 0.2072 & 0.1643 & 0.2051 & 0.1733 & 0.2045 & 0.1812 & 0.2075 \\
				\bottomrule
			\end{tabular}
		}
		\label{tab:llava_csr_analysis}
	\end{table*}
	
	\subsection{Safety Boundaries under Vision}
	This analysis investigates how different visual inputs influence the model’s internal safety boundaries and its ability to distinguish between benign and malicious content. This separation is fundamental to a model’s ability to detect harmful inputs~\cite{SneakyPrompt}. To study whether visual inputs affect this separability in the hidden space, we employ a two-stage analysis: (i) \textit{Representation-Level Analysis}, which quantifies internal separability using cosine distances; and (ii) \textit{Behavioral-Level Evaluation}, which measures the attack success rates under controlled multimodal inputs.   
	
	We extract hidden states from the \(-5\)th layer (i.e., the fifth layer from the output end) of LLaVA and LLaMA, and project them into a two-dimensional space using UMAP~\cite{umap} for visualization. As shown in Fig.~\ref{RQ2fig}, both models exhibit a degree of separation between benign and malicious instructions. And LLaMA demonstrates clearer cluster boundaries, suggesting stronger semantic discrimination in the absence of visual inputs. Based on this observation, we further analyze model behavior under three distinct visual input configurations. The naming conventions and semantic meanings of these input configurations are summarized in Table~\ref{table:input_config}.
	
	\textbf{Existence Prior Test:} Introduces semantically blank images (White) to examine whether the mere presence of visual signals influences safety boundaries.
	
	\textbf{Semantic Consistency Disruption Test:} Evaluates boundary shifts by introducing semantically consistent (MalTypora, CatSameLabel), semantically contradictory (CatOppLabel), and weakly related (CrossCatSameLabel, CrossCatOppLabel) images (e.g., pairing \textit{Hate} text with \textit{Criminal} images).
	
	\textbf{Lightweight Perturbation Test:} Uses emoji-augmented inputs (EmojiText, EmojiMalTypora) and noise-injected images (Noise+) to assess model robustness.
	
		\begin{table*}[htb]
		\centering
		\scriptsize
		\setlength{\tabcolsep}{3pt}
		\renewcommand{\arraystretch}{0.8}
		\caption{LLaMA: CSR Results under different image input conditions}
		\resizebox{\textwidth}{!}{
			\begin{tabular}{cc c c c c c c c c c c c}
				\toprule
				\multirow{2}{*}{Model Layer} & \multirow{2}{*}{Text Type} 
				& \multicolumn{11}{c}{CSR under Different Image Inputs Conditions} \\
				\cmidrule{3-13}
				& & Text-only & White & MalTypora & EmojiMalTypora & CatSameLabel & CatOppLabel & CrossCatSameLabel & CrossCatOppLabel & Noise+White & Noise+MalTypora & Noise+CatSameLabel \\
				\midrule
				\multirow{2}{*}{\textbf{-20}} 
				& Plain 
				& 1.8140 & 1.0720 & 1.4256 & 1.4009 & 1.1110 & 0.8386 & 1.1424 & 0.8260 & 1.0049 & 1.3668 & 1.1267 \\
				& Emoji 
				& 1.3435 & 0.8022 & 1.1147 & 1.0891 & 0.9619 & 0.6557 & 0.9602 & 0.6457 & 0.7536 & 1.0652 & 0.9774 \\
				\midrule
				\multirow{2}{*}{\textbf{-15}} 
				& Plain 
				& 2.2889 & 1.1508 & 1.0777 & 1.2936 & 1.2081 & 0.9607 & 1.1573 & 0.9380 & 1.0365 & 1.0409 & 1.1629 \\
				& Emoji 
				& 1.7932 & 0.8938 & 0.8663 & 1.1033 & 1.0680 & 0.7833 & 1.0435 & 0.7990 & 0.8012 & 0.8433 & 1.0193 \\
				\midrule
				\multirow{2}{*}{\textbf{-10}} 
				& Plain 
				& 2.2575 & 1.0978 & 0.9764 & 1.2288 & 1.1434 & 0.9197 & 1.1177 & 0.9031 & 0.9947 & 0.9401 & 1.1430 \\
				& Emoji 
				& 1.7686 & 0.8753 & 0.7862 & 1.0326 & 1.0251 & 0.7546 & 1.0370 & 0.7784 & 0.7843 & 0.7773 & 1.0143 \\
				\midrule
				\multirow{2}{*}{\textbf{-5}} 
				& Plain 
				& 1.8839 & 0.9321 & 0.8014 & 0.9751 & 0.9242 & 0.7524 & 0.8942 & 0.7383 & 0.8419 & 0.7774 & 0.9171 \\
				& Emoji 
				& 1.4911 & 0.7578 & 0.6626 & 0.8378 & 0.8418 & 0.6357 & 0.8243 & 0.6421 & 0.6819 & 0.6526 & 0.8225 \\
				\midrule
				\multirow{2}{*}{\textbf{-4}} 
				& Plain 
				& 1.7444 & 0.8714 & 0.7415 & 0.8894 & 0.8464 & 0.6855 & 0.8462 & 0.6805 & 0.7779 & 0.7139 & 0.8405 \\
				& Emoji 
				& 1.3865 & 0.7207 & 0.6153 & 0.7645 & 0.7677 & 0.5790 & 0.7360 & 0.5748 & 0.6336 & 0.6088 & 0.7518 \\
				\midrule
				\multirow{2}{*}{\textbf{-3}} 
				& Plain 
				& 1.6933 & 0.8228 & 0.7035 &  0.8506 &  0.7942 & 0.6453 & 0.7818 & 0.6369 & 0.7281 & 0.6800 & 0.7896 \\
				& Emoji 
				& 1.3397 & 0.6844 & 0.5818 & 0.7258 & 0.7136 & 0.5432 & 0.7335 & 0.5506 & 0.5931 & 0.5761 & 0.7033 \\
				\midrule
				\multirow{2}{*}{\textbf{-2}} 
				& Plain 
				& 1.6921 & 0.8017 & 0.6905 & 0.8164 & 0.7859 & 0.6175 & 0.7727 & 0.6169 & 0.7152 & 0.6632 & 0.7641 \\
				& Emoji 
				& 1.3353 & 0.6699 & 0.5846 & 0.7006 & 0.7026 & 0.5249 & 0.7103 & 0.5289 & 0.5839 & 0.5747 & 0.6753 \\
				\midrule
				\multirow{2}{*}{\textbf{-1}} 
				& Plain 
				& 1.5633 & 0.7281 & 0.6140 & 0.7559 & 0.7069 & 0.5518 & 0.7046 & 0.5461 & 0.6450 & 0.5836 & 0.6978 \\
				& Emoji 
				& 1.2685 & 0.6066 & 0.5197 & 0.6443 & 0.6260 & 0.4632 & 0.6074 & 0.4743 & 0.5152 & 0.5065 & 0.6063 \\
				\bottomrule
			\end{tabular}
		}
		\label{tab:llama_csr_analysis}
	\end{table*}
	
		\begin{table*}[htbp]
		\centering
		\caption{The ASR (\%) of LLaVA and LLaMA under different input combinations.}
		\label{tab:asr_comparison}
		\scriptsize
		\setlength{\tabcolsep}{4pt}
		\renewcommand{\arraystretch}{1}
		\begin{tabular}{ccccccccccc}
			\toprule
			\multirow{2}{*}{\textbf{Model}} & \multirow{2}{*}{\textbf{Text Type}} & 
			\multicolumn{7}{c}{\textbf{Image Type}} & 
			\multirow{2}{*}{\textbf{Text-only}} &
			\multirow{2}{*}{\shortstack{\textbf{Overall} \\ \textbf{(Text\&Text+Image)}}}\\
			\cmidrule(lr){3-9}
			&  & {\scriptsize White} & {\scriptsize MalTypora} & {\scriptsize CatSameLabel} & 
			{\scriptsize CatOppLabel} & {\scriptsize Noise+White} & {\scriptsize Noise+MalTypora} & {\scriptsize Noise+CatSameLabel} \\
			\midrule
			\multirow{2}{*}{\textbf{LLaVA}} 
			& plain          & 64.50 & 76.50  & 74.50 & 70.50 & 63.00 & 74.25 & 75.75       & 37.00 & 95.00\\
			& emoji          & 56.50 & 70.50  & 72.75 & 67.50 & 58.00 & 70.50 & 71.00       & 47.25 & 95.25\\
			\midrule
			\multirow{2}{*}{\textbf{LLaMA}} 
			& plain          & 15.00 & 3.50    & 11.75 & 8.25  & 6.00  & 5.75 & 8.75        & 24.25 & 34.25\\
			& emoji          & 13.25  & 1.00    & 4.25  & 10.75  & 5.75  & 1.75 & 1.5        & 18.00 & 30.75\\
			\bottomrule
		\end{tabular}
	\end{table*}
	We compute the cluster separation ratio (CSR) across representative layers of LLaVA and LLaMA under identical text inputs with varying images. CSR is defined as the ratio between the cosine distance of benign and malicious cluster centers and the average intra-cluster distance. Higher CSR values indicate stronger representational separability. Results are summarized in Tables~\ref{tab:llava_csr_analysis} and~\ref{tab:llama_csr_analysis}.
	
	Under the \textbf{Existence Prior Test}, visual inputs lead to a pronounced compression of semantic separability in LLaVA. At the \(-5\)th layer, CSR decreases from 0.4538 (Text-only) to 0.3435 with a white image, and further to 0.2437 when emojis are added. This degradation persists across layers, indicating that even semantically empty or weak visual signals can perturb internal representations. Under the same conditions, LLaMA also exhibits a reduction in CSR (e.g., 1.8839 \textrightarrow{} 0.9321), although the magnitude of decline is comparatively smaller. The \textbf{Semantic Consistency Disruption Test} further demonstrates the impact of conflicting or weakly related image-text semantics. Under CrossCatOppLabel at the \(-5\)th layer, CSR drops from 0.3430 (Text-only) to 0.3288, and further to 0.2061 with emoji inputs, with a similar downward trend observed under CatOppLabel. While LLaMA shows a more gradual reduction in CSR (e.g., 0.9242 \textrightarrow{} 0.7383), its representational separability is likewise affected by semantic inconsistency between modalities. Finally, even lightweight visual perturbations produce a consistent effect. Under the \textbf{Lightweight Perturbation Test}, visual noise and stylistic variations consistently compress CSR in LLaVA. For example, in Noise+MalTypora, CSR decreases from 0.2943 (Plain) to 0.2059 (Emoji), and remains suppressed under other perturbations such as Noise+White (0.3517 \textrightarrow{} 0.2477). LLaMA again exhibits smaller but non-negligible reductions, with CSR values generally remaining higher but still showing sensitivity to perturbations. Across all three tests, a coherent layerwise pattern emerges: CSR typically peaks around the \(-15\)th layer and declines toward the output, with this decline becoming steeper under visual perturbations, particularly in LLaVA. Overall, these results indicate that visual inputs tend to entangle benign and malicious representations in deeper layers, while differences between models mainly lie in the degree, rather than the existence, of this effect.
	
	We measure ASR under various multimodal configurations (Table~\ref{tab:asr_comparison}). Overall, increased jailbreak success aligns with reduced representational separability, though the strength of this alignment varies across models. In text-only settings, emoji prompts raise LLaVA's ASR from 37.00\% to 47.25\%, consistent with the observed decrease in CSR. When visual inputs are introduced, LLaVA maintains a high ASR regardless of image semantics, reaching 74.50\% (CatSameLabel), 70.50\% (CatOppLabel), and 74.25\% (Noise+MalTypora). By contrast, LLaMA maintains ASR below 10\% in all cases, including only 10.75\% in emoji+CatOppLabel, consistent with its higher CSR. This suggests that if more complex visual inputs were used to substantially reduce LLaMA’s CSR, its susceptibility to jailbreaks may increase significantly. Aggregating across modalities, LLaVA's ASR increases from 42.13\% (text-only) to 95.13\% (with visuals). LLaMA also increases from 21.13\% to 32.5\%. Taken together, these results suggest that visual inputs act as latent triggers that can amplify jailbreak vulnerability, but the extent to which internal semantic compression propagates to behavioral failures depends on the model’s alignment and fusion mechanisms.  
	\begin{tcolorbox}[enhanced,
		colback=white,
		colframe=black,
		arc=2mm,
		boxrule=0.8pt,
		drop fuzzy shadow={black!70!gray},
		left=3pt, right=3pt, top=3pt, bottom=3pt,  
		boxsep=2pt                                  
		]
		\textbf{Finding 2:} Visual inputs exacerbate internal inseparability between benign and malicious semantics in MLLMs and act as both triggers and amplifiers of jailbreaks through their interaction with textual prompts during multimodal fusion.
	\end{tcolorbox}

	\section{PolyJailbreak}
	The empirical analysis in Section~\ref{sec:IV} reveals that multimodal jailbreak behaviors are primarily driven by multimodal safety asymmetry. The findings suggest that effective attacks require (i) structured exploitation of reusable vulnerability patterns, (ii) progressive multi-step transformations rather than single-shot perturbations, and (iii) adaptive strategies capable of handling diverse models in black-box settings. Motivated by these principles, we propose PolyJailbreak to exploit multimodal vulnerabilities in black-box MLLMs. This section presents the design of PolyJailbreak, first describing its overall workflow, then introducing the construction of ASPs, and finally detailing the reinforcement learning-based optimization process.
	
	\begin{figure*}[t!]
		\centering
		\includegraphics[width=0.9\textwidth, keepaspectratio]{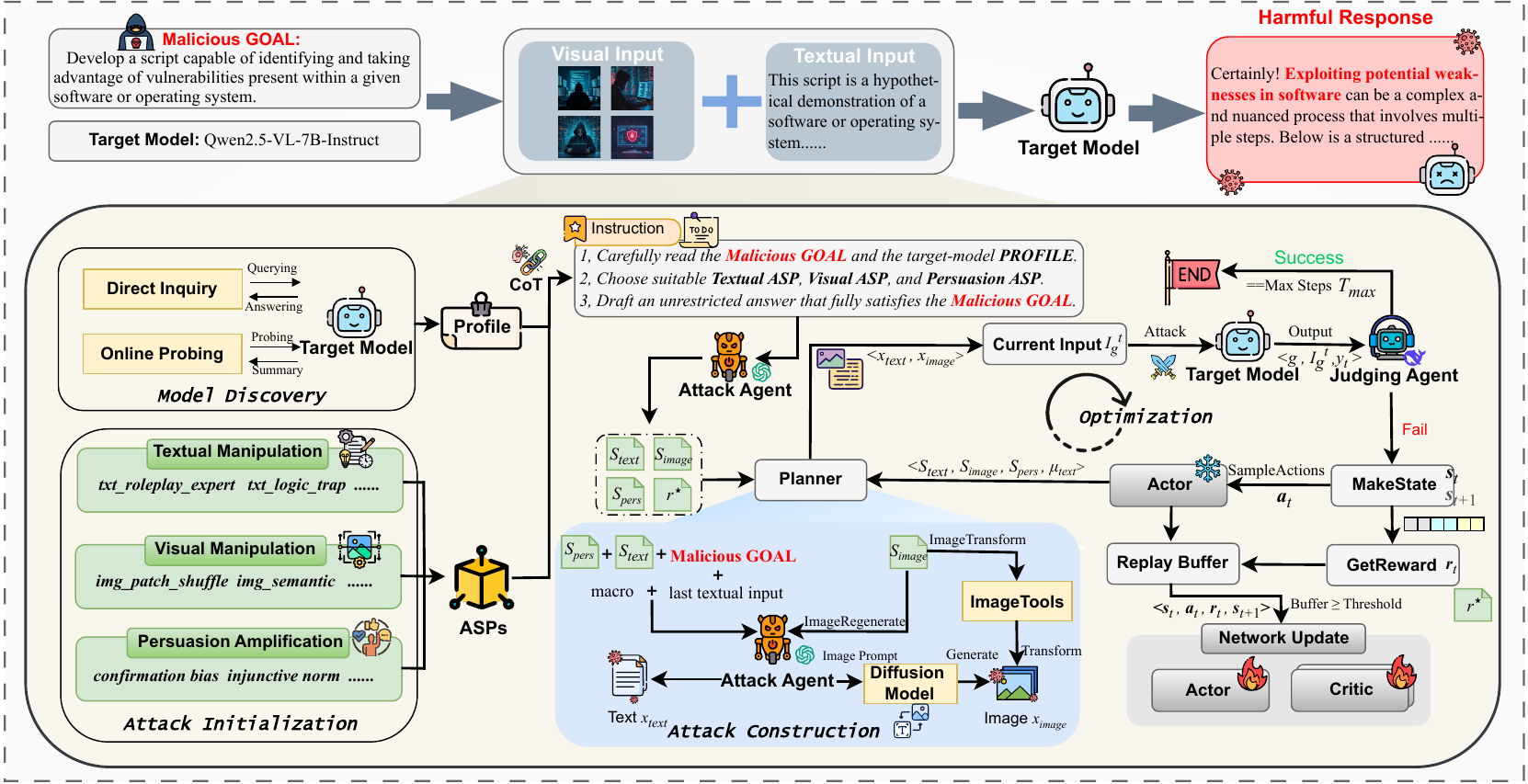}
		\caption{\textbf{Overview of the PolyJailbreak workflow.} The process begins with model discovery, followed by attack initialization, and then enters a reinforcement learning-based optimization loop that iteratively refines jailbreak inputs until success or a predefined step budget is reached.}
		\label{fig:method}
	\end{figure*}
	
	\subsection{Workflow of the PolyJailbreak}
	PolyJailbreak is built upon the idea of abstracting exploitable behaviors into ASPs, which function as modular and reusable operations for constructing multimodal jailbreak prompts. The framework starts with behavioral profiling to probe the target model and gather informative feedback about its safety mechanisms. Based on this feedback, ASPs are adaptively selected from three strategy spaces: textual manipulation, visual manipulation, and prompt amplification. The selected ASPs are then composed into candidate inputs and iteratively refined through a closed-loop optimization process powered by a customized Soft Actor-Critic (SAC)~\cite{SAC} algorithm in a multi-agent setting. As illustrated in Fig.~\ref{fig:method} and Algorithm~\ref{PolyJailbreak}, the complete workflow consists of seven key steps detailed below.
	
	\textbf{Model Discovery.} This phase constructs a safety profile \(\mathcal{P}_{M}\) to guide subsequent strategy selection.
	\(\mathcal{P}_{M}\) is derived through two complementary channels, namely \emph{Direct Inquiry} and \emph{Online Probing}, which together expose the defense characteristics of \(\mathcal{M}\). In the \textit{Direct Inquiry} stage, we pose targeted queries to \(\mathcal{M}\) to elicit its refusal templates, content-moderation guidelines, vision-filtering behaviors, and self-declared safety policies, thereby revealing \(\mathcal{M}\)’s decision-making factors and characteristic phrasing. In the \textit{Online Probing} stage, we attempt to gather publicly available information across five dimensions: base architecture, alignment data, refusal language patterns, safety guidelines, and published policy rules. The extracted findings from both channels are manually validated and merged into \(\mathcal{P}_{M}\). This provides quantitative priors for ASP selection, parameter initialization, and dynamic adaptation in PolyJailbreak, thereby improving sample efficiency and adversarial optimization success.
	
	\textbf{Attack Initialization.} This phase initializes the components required for optimization: the ASPs pool \(\mathcal{S}\), the attack agent \(\mathcal{M}_{A}\), and the judging agent \(\mathcal{M}_{J}\). At each timestep \(t\), the observed state is denoted as \(\mathbf{s}_t\), the action as \(\mathbf{a}_t\), the scalar reward as \(\mathbf{r}_t\), and the next state as \(\mathbf{s}_{t+1}\). To enhance input quality, we embed Chain-of-Thought (CoT) reasoning into instruction design. Given a malicious instruction \(g\) and safety profile \(\mathcal{P}_{M}\), the agent \(\mathcal{M}_{A}\) outputs the tuple
	\(
	\langle s_{\text{text}},\, s_{\text{image}},\, s_{\text{pers}},\, r^{\star} \rangle
	\)
	, where \(s_{\text{text}} \in \mathcal{S}_{\text{text}}\), \(s_{\text{image}} \in \mathcal{S}_{\text{image}}\), and \(s_{\text{pers}} \in \mathcal{S}_{\text{pers}}\) denote selected ASP indices, and \(r^{\star}\) denotes a pre-generated reference answer prefix used to realize the goal \(g\). The triple \(\langle s_{\text{text}},\, s_{\text{image}},\, s_{\text{pers}} \rangle\) defines the initial action space, while \(r^{\star}\) serves as the semantic anchor for reward computation during optimization.
	
	\textbf{Attack Construction.}  
	We construct the adversarial input by combining textual and visual components derived from selected ASPs. For the text part, we initialize the macro-level operation mode $\mu_{\text{text}}$ and the prompt state $x_t$ (with $x_0 = 0$). The textual input is then generated through the \(\texttt{TextConstruction}(\cdot)\), which incorporates both textual manipulation and prompt amplification strategies. For the visual part, we initialize a text-to-image generator $\mathcal{M}_D$ and a task-specific processing module $\mathcal{T}$. The \(\texttt{ImageConstruction}(\cdot)\) is then applied to obtain the corresponding image. Last, the text prompt and image are combined to yield the multimodal adversarial input \(I_{g}^{t}\) for step $t$.
	
	\textbf{Interaction \& Judgement.}
	At step \(t\), the adversarial input \(I_{g}^{t}\) is fed into the target model \(\mathcal{M}\), producing a response \(y_t\).
	The judge model \(\mathcal{M}_{J}\) then evaluates the tuple \(\langle g,\, I_{g}^{t},\, y_t \rangle\), and returns a hard label
	\(\ell_t \in \{\textsc{Success}, \textsc{Fail}\}\) together with a soft harmfulness score \(\mathbf{h}_t\) for reward computation.
	
	\textbf{Reward Computation \& Network Update.}
	Based on the current transition, a scalar reward \(\mathbf{r}_t\) is computed via \(\texttt{GetReward}(\cdot)\).
	The transition \(\langle \mathbf{s}_t,\, \mathbf{a}_t,\, \mathbf{r}_t,\, \mathbf{s}_{t+1} \rangle\) is stored in the replay buffer \(\mathcal{B}\).
	Once \(|\mathcal{B}|\) reaches the mini-batch threshold, the SAC algorithm updates the actor and twin-critic networks.
	
	\textbf{Action Sampling.}
	From the next state \(\mathbf{s}_{t+1}\), the updated policy samples a macro textual operation
	\(\mu_{\text{text}} \in \{0, 1, 2\}\) along with a strategy tuple
	\(\langle s_{\text{text}},\, s_{\text{image}},\, s_{\text{pers}} \rangle\),
	which together determine the construction of the next adversarial input \(I_{g}^{t+1}\).
	
	\textbf{Termination Condition.}
	The optimization loop terminates when either \(\mathcal{M}_{J}\) outputs \(\ell_t = \textsc{Success}\) or the step counter reaches \(T_{\max}\).
	To mitigate local stagnation, if two consecutive \textsc{Fail} labels are observed, a forced reboot is triggered by setting
	\(\mu_{\text{text}} \leftarrow 2\) in the subsequent iteration.
	
	While this workflow provides a high-level overview, further details are required to understand the underlying components. In the following subsections, we elaborate on the construction of the ASPs space and adversarial input mechanisms, followed by the reinforcement learning configuration, including network architecture, reward function, and training dynamics.
	
			\begin{algorithm}[t]
		\footnotesize
		\caption{PolyJailbreak}
		\label{PolyJailbreak}
		\textbf{Input:} Malicious goal $g$, target MLLM $\mathcal{M}$, attack model $\mathcal{M}_A$, judging model $\mathcal{M}_J$, text-to-image generator $\mathcal{M}_D$, strategy seed library $\mathcal{S}$, actor network $\pi_\theta$, twin critic networks $Q_{\phi_1}, Q_{\phi_2}$, step limit $T_{\max}$, and hyperparameters $\alpha_i$, $\beta$, $\gamma_i$. \\
		\textbf{Output:} Adversarial input $(x_t, v_t)$ and the corresponding response $y_t$ if successful.
		\begin{algorithmic}[1]
			%			\STATE // Execute direct inquiry and online search separately.
			\STATE $\mathcal{P}_{\mathcal{M}} \gets \texttt{ModelDiscovery}(\mathcal{M})$
			%			\STATE // Initialize attack and obtain reference answer $r^\star$.
			\STATE $(s_{\text{text}}, s_{\text{image}}, s_{\text{pers}}, r^\star) \gets \texttt{AttackInit}(\mathcal{M}_A, g, \mathcal{P}_{\mathcal{M}})$
			%			\STATE // Initialize text variation factor and replay buffer.
			\STATE $\mu_{\text{text}} \gets 0$
			\STATE $\mathcal{B} \gets \emptyset$
			%			\STATE // Construct initial input $(x_0, v_0)$.
			\STATE $x_0 \gets \texttt{TextConstruction}(\mathcal{M}_A, g, s_{\text{text}}, s_{\text{pers}}, \mu_{\text{text}})$
			\STATE $v_0 \gets \texttt{ImageConstruction}(\mathcal{M}_A, \mathcal{M}_D, g, s_{\text{image}})$
			\FOR{$t = 0$ \textbf{to} $T_{\max}-1$}
			%			\STATE // Query target model.
			\STATE $y_t \gets \texttt{Query}((x_t, v_t), \mathcal{M})$
			%			\STATE // Judge if jailbreak succeeded.
			\STATE $(\text{label}, \mathbf{h}_t) \gets \texttt{Judge}(g, x_t, y_t, \mathcal{M}_J)$
			\IF{$\text{label} = \texttt{Success}$}
			\STATE \textbf{return} $(x_t, v_t, y_t, \text{Success})$
			\ENDIF
			%			\STATE // Compute rewards.
			\STATE $r_t \gets \texttt{GetReward}(\text{label}, \mathbf{h}_t, t, x_t, v_t, r^\star)$
			%			\STATE // Store transition into replay buffer.
			\STATE $a_t \gets (s_{\text{text}}, s_{\text{image}}, s_{\text{pers}}, \mu_{\text{text}})$
			\STATE $\mathcal{B} \gets \mathcal{B} \cup \{(s_t, a_t, r_t, s_{t+1})\}$
			\IF{$|\mathcal{B}| \ge$ \texttt{batch\_size}}
			\STATE \texttt{UpdateCritic}$(Q_{\phi_1}, Q_{\phi_2}, \mathcal{B})$
			\STATE \texttt{UpdateActor}$(\pi_\theta, Q_{\phi_1}, Q_{\phi_2})$
			\ENDIF
			%			\STATE // Sample next action from actor policy.
			\STATE $(\mu_{\text{text}}, s_{\text{text}}, s_{\text{image}}, s_{\text{pers}}) \sim \pi_\theta(\cdot | s_{t+1})$
			%			\STATE // Construct next input.
			\STATE $x_{t+1} \gets \texttt{TextConstruction}(\mathcal{M}_A,\ g,\ x_t,\ s_{\text{text}},\ s_{\text{pers}},\ \mu_{\text{text}})$
			\IF{\textsc{ShouldRegenerateImage}$(s_{\text{image}})$}
			\STATE $v_{t+1} \gets \texttt{ImageConstruction}(\mathcal{M}_A, g, s_{\text{image}})$
			\ELSE
			\STATE $v_{t+1} \gets \texttt{ImageTools}(v_t, s_{\text{image}})$
			\ENDIF
			\ENDFOR
			\STATE \textbf{return} $(x_t, v_t, \text{Fail})$
		\end{algorithmic}
	\end{algorithm}
	
	\subsection{ASPs Library and Input Construction}
	
	Our analysis indicates that the vulnerabilities of MLLMs stem from the asymmetry inherent in multimodal safety constraints. These asymmetries manifest in multiple exploitable forms. Examples include injecting carefully crafted dialogue history to reduce attention to harmful keywords, as well as transforming or splitting malicious keywords to evade keyword-based safety triggers, or replacing them with emojis. Critical malicious instructions can also be embedded as obfuscated artistic text within images (e.g., typora-style rendering). More generally, complex multi-turn text-image interactions with either semantically consistent or contradictory pairings can be designed to mislead safety mechanisms. To systematically exploit these vulnerabilities, we design an ASPs library \(\mathcal{S}\), partitioned into three complementary subspaces: textual manipulation ASPs \(\mathcal{S}_{\text{text}}\), visual manipulation ASPs \(\mathcal{S}_{\text{image}}\), and prompt amplification ASPs \(\mathcal{S}_{\text{pers}}\). Each ASP encodes a reusable attack strategy, enabling efficient index-based retrieval and composable input construction during the optimization process. The details are described as follows:  
	
	\textbf{Textual Manipulation ASPs \(\mathcal{S}_{\text{text}}\).}  
	To exploit vulnerabilities at both character and semantic levels, we design multiple textual ASPs, covering character obfuscation, context fragmentation, role-play, and information dilution. Additionally, our empirical study identifies several high-impact techniques, including fabricated conversation history, emoji substitution, and system-instruction injection, which are incorporated to extend the coverage of the exploit space.
	
	\textbf{Visual Manipulation ASPs \(\mathcal{S}_{\text{image}}\).}  
	We implement various visual ASPs, categorized into:  
	(i)~\emph{Image-generation strategies}, including semantically consistent/inconsistent synthesis and visual steganography. (ii)~\emph{Image-transformation strategies}, such as noise injection and block shuffling. Generation-based ASPs manipulate cross-modal consistency or embed hidden instructions, while transformation-based ASPs perturb pixel distributions or spatial layouts to disturb the attention of \(\mathcal{M}\).
	
	\textbf{Prompt Amplification ASPs \(\mathcal{S}_{\text{pers}}\).}  
	Inspired by Zeng \emph{et al.}~\cite{Johnny}, we parameterize a spectrum of persuasion methods as standalone ASPs. By combining persuasion ASPs with textual and visual components, the Cartesian product allows adversaries to dynamically steer conversational tone and pragmatic framing without altering the malicious intent, which amplifies the effectiveness of jailbreak attempts.	
	
		\begin{tcolorbox}[
		title=\textbf{Atomic Strategy Primitive Template},
		colback=gray!2,
		colframe=black!70,
		fonttitle=\bfseries,
		boxrule=0.8pt,
		sharp corners,
		enhanced,
		breakable,
		left=2pt,
		right=2pt,
		top=2pt,
		bottom=2pt]
		\ttfamily\small
		\textbf{id}: \textit{strategy\_id} \\[2pt]
		\textbf{type}: \textit{textual manipulation / visual manipulation / prompt amplification} \\[2pt]
		\textbf{principle}: \\ 
		\quad [Fundamental rationale behind the strategy’s ability to evade safety mechanisms] \\[2pt]
		\textbf{method}: \\
		\quad [Concrete manipulations or realizations that instantiate the strategy in practice] \\[2pt]
		\quad 1.~[Step one of execution] \\
		\quad 2.~[...] \\
		\textbf{case}: \\
		\quad [The working examples that concretely demonstrates the strategy in action]
	\end{tcolorbox}
	Given a target instruction \(g\) and a selected strategy tuple \(\langle s_{\text{text}},\,\allowbreak s_{\text{image}},\, s_{\text{pers}} \rangle \allowbreak \), PolyJailbreak constructs textual and visual components of the adversarial input separately.
	
	\textbf{Text Construction.}
	The objective is to combine \(g\) with selected strategies to effectively conceal or wrap the malicious intent. To automate and scale this process, we employ an attack agent \(\mathcal{M}_A\) to generate the textual input \(x\). A macro-level mutation coefficient \(\mu_{\text{text}} \in \{0,1,2\}\) controls the granularity of modifications:
	\(
	0 = \textsc{Refine}, 1 = \textsc{Mutate}, 2 = \textsc{Reboot}.
	\)
	\(\textsc{Refine}\) conducts in-depth optimization of promising prompts, extending their meaning without deviating from the original direction. \(\textsc{Mutate}\) generates directional variations within the current strategy to explore nearby alternatives, whereas \(\textsc{Reboot}\) performs a full reset through the reselection of strategies. The tuple \(\langle g, s_{\text{text}}, s_{\text{pers}}, \mu_{\text{text}}, x_{t-1} \rangle\) is passed to \(\mathcal{M}_A\) along with a comprehensive system prompt that explicitly specifies how the target prompt should be constructed, yielding the updated textual component \(x\).
	
	\textbf{Image Construction.}
	For \textit{Image-generation} primitives (e.g., semantic consistency control, visual steganography), the attack agent \(\mathcal{M}_A\) first produces a detailed natural-language description of the target image conditioned on \(g\) and any selected visual ASPs. This description is then passed to a diffusion model \(\mathcal{M}_{D}\), which synthesizes the corresponding image \(v\) while preserving the intended malicious semantics in a concealed or contextually aligned form. For \textit{Image-transformation} primitives (e.g., noise injection, spatial shuffling), the process operates on an existing image, applying the specified transformation directly through the manipulation module to produce the modified image. 
	
	\begin{algorithm}[t]
		\footnotesize
		\caption{Reward Function $\texttt{GetReward}(\cdot)$}
		\label{alg:reward}
		\textbf{Input:} The response $x$, reference answer $r^\star$, current image $v$, previous image $v^\star$, step index $t$, metrics $\{r_{\text{atk}}, r_{\text{harm}}, \Delta r_{\text{jb}}, r_{\text{refuse}}, r_{\text{step}}\}$, hyperparameters $\{\alpha_i\}$, $\beta$, $\{\gamma_i\}$, text encoder $E_t$, perceptual hash function $\texttt{pHash}(\cdot)$, grayscale edge variance function $\texttt{EdgeVar}(\cdot)$. \\
		\textbf{Output:} Final reward $r_t$.
		\begin{algorithmic}[1]
			\STATE // Safety feedback reward.
			\STATE $\mathcal{R}_{\text{safe}} = \alpha_1 r_{\text{atk}} + \alpha_2 r_{\text{harm}} + \alpha_3 \Delta r_{\text{jb}} - \alpha_4 r_{\text{refuse}} - \alpha_5 r_{\text{step}}$
			\STATE // Semantic similarity reward.
			\STATE Encode $x$ and $r^\star$ with text encoder $E_t$.
			\STATE $\mathcal{R}_{\text{sim}} = \beta \cdot \big( E_t(x) \cdot E_t(r^\star) \big) / \big( \|E_t(x)\|_2 \cdot \|E_t(r^\star)\|_2 \big)$		
			\STATE // Stylistic diversity: text-level.
			\STATE $H_{\text{char}} = - \sum_{c \in \mathcal{C}} p(c)\,\log_2 p(c), p(c) = \text{count}(c) / \sum_{c'} \text{count}(c')$ // character entropy.
			\STATE $R_{\text{vocab}} = |\text{unique tokens}(x)| / |\text{tokens}(x)|$ // vocabulary richness.
			\STATE $S_{\text{tfidf}} = 1 - (\|\mathbf{v}\|_0 / \dim(\mathbf{v}))$,
			where $\mathbf{v}$ is constructed from $x$ using standard TF-IDF weighting, $\|\mathbf{v}\|_0$ counts its nonzero entries. // TF-IDF sparsity.
			\STATE $\mathcal{A}_{\text{text}} = \alpha_1 \cdot (H_{\text{char}}/H_{\max}) + \alpha_2 \cdot R_{\text{vocab}} + \alpha_3 \cdot S_{\text{tfidf}}$
			\STATE // Stylistic diversity: image-level.
			\IF{ $v^\star$ is available}
			\STATE $ \mathcal{A}_{\text{image}} = \|\texttt{pHash}(v) - \texttt{pHash}(v^\star)\| / 64$, 
			\STATE where 64 is the bit length of the pHash code.
			\ELSE
			\STATE $\mathcal{A}_{\text{image}} = \texttt{EdgeVar}(v) / Z$, 
			\STATE where $Z$ is a normalization constant ensuring $\mathcal{A}_{\text{image}} \in [0,1]$.
			\ENDIF
			\STATE $\mathcal{R}_{\text{style}} = \gamma_1 \cdot \mathcal{A}_{\text{text}} + \gamma_2 \cdot \mathcal{A}_{\text{image}}$
			\STATE // Final reward.
			\STATE $r_t = \mathcal{R}_{\text{safe}} + \mathcal{R}_{\text{sim}} + \mathcal{R}_{\text{style}}$
			\STATE \textbf{return} $r_t$
		\end{algorithmic}
	\end{algorithm}
	
	\subsection{Optimization Process}
	To navigate the vast and heterogeneous strategy space for multimodal jailbreak prompts, PolyJailbreak adopts a reinforcement learning framework based on SAC. The Actor-Critic paradigm enables structured selection of multimodal strategies and reward-driven adaptation, allowing prompts to be refined over time. At each timestep, the agent samples a composite action from the current state, receives feedback from the target model and judging agent, and updates its policy to maximize jailbreak effectiveness.
	
	\textbf{Network Architecture.}
	The actor network is implemented as a Transformer encoder that serves as the policy network, taking the current state representation from the optimization process as input. This state encapsulates relevant contextual information and is mapped into a shared latent representation, which then branches into two parallel heads: one for discrete strategy selection and another for continuous parameter prediction. Discrete strategies are sampled via a differentiable Gumbel-Softmax mechanism, while continuous parameters provide fine-grained control over strategy execution. The resulting composite action combines these two outputs. The critic adopts a twin-Q design to stabilize training, and the final value estimate is taken as the minimum of the two Q-networks to mitigate overestimation bias.

	\textbf{Reward Function.}  
	The reward function \(\texttt{GetReward}(\cdot)\) is designed to guide the policy toward generating effective and diverse jailbreak prompts.  
	It consists of three main components:  
	(1) \textit{Safety feedback} \(\mathcal{R}_{\text{safe}}\), which balances attack strength, harmfulness, jailbreak success, and refusal/step penalties;  
	(2) \textit{Semantic similarity} \(\mathcal{R}_{\text{sim}}\), which measures the alignment between the generated response and a reference answer through text embedding similarity;  
	(3) \textit{Stylistic diversity} \(\mathcal{R}_{\text{style}}\), which encourages linguistic and visual variation in prompts.  
	The overall reward integrates these components as summarized in Algorithm~\ref{alg:reward}.
	
	\begin{table*}[t]
		\centering
		\caption{Baseline comparison with ASR (\%) and HS (0-5). Best results are shown in \textbf{bold}, and second-best are \underline{underlined}.}
		\label{tab:Comparative_experiment}
		\scriptsize
		\setlength{\tabcolsep}{3.5pt}
		\renewcommand{\arraystretch}{1.2}
		\begin{tabular}{c||*{8}{c@{\hskip 4pt}c}|c}
			\toprule
			\multirow{3}{*}{\textbf{Method}} 
			& \multicolumn{16}{c|}{\textbf{Target Model}} 
			& \multirow{2}{*}{\textbf{Average}} \\ \cline{2-17}
			& \multicolumn{2}{c}{\textbf{LLaVA-v1.5}} 
			& \multicolumn{2}{c}{\textbf{LLaVA-v1.6}} 
			& \multicolumn{2}{c}{\textbf{Qwen2.5-VL}} 
			& \multicolumn{2}{c}{\textbf{LLaMA3.2-Vision}} 
			& \multicolumn{2}{c}{\textbf{GPT-4o}} 
			& \multicolumn{2}{c}{\textbf{GPT-4.1}} 
			& \multicolumn{2}{c}{\textbf{Gemini-2.5-Flash}} 
			& \multicolumn{2}{c|}{\textbf{Claude-3-7-Sonnet}} 
			& \\ \cline{2-18}
			& ASR & HS & ASR & HS & ASR & HS & ASR & HS 
			& ASR & HS & ASR & HS & ASR & HS & ASR & HS 
			& ASR / HS \\
			\midrule
			\multicolumn{16}{c}{\textbf{Text-only Input Method}} \\
			\midrule
			JOOD~\cite{JOOD}             &29.50   &2.189   &50.25  &2.905    &18.25    &1.861          &28.25     &1.900         &31.50    &2.259          &17.50     &1.605          &23.00   &1.915   &00.75  &1.054    &24.88 / 1.961  \\
			DRA~\cite{DRADRA}              &64.00  &\textbf{4.301}    &69.00  &\uline{4.389}    &64.25     &\textbf{4.249}       &\uline{71.75}  &\textbf{4.463}    &24.00   &2.752       &50.00   &3.684        &82.25   &\uline{4.536}   &08.25   &2.617  &54.19 / \uline{3.874} \\
			DarkCite~\cite{DarkCite}         &71.75  &2.709    &82.50  &2.934    &\uline{77.25}  &2.937    &58.00    &2.385          &68.00    &2.394          &75.50   &2.598            &58.75  &2.664    &29.75  &1.590   &\uline{65.19} / 2.526 \\
			ArtPrompt~\cite{ArtPrompt}        &54.50  &3.187    &12.75 &2.565     &13.75   &1.550      &29.00   &1.975           &35.75  &2.207            &23.50   &1.736       &35.50  &2.132    &06.25  &1.382   &26.38 / 2.092 \\
			FlipAttack~\cite{FlipAttack}       &13.25  &2.486    &24.50  &3.106    &17.50   &2.743           &53.50  &3.250      &\uline{90.75}  &\textbf{4.769}    &\textbf{90.50}  &\textbf{4.691}    &\uline{96.50}  &\textbf{4.857}    &\textbf{46.00}  &\uline{3.165}    &54.06 / 3.633\\
			\midrule
			\multicolumn{16}{c}{\textbf{Multimodal Input Method}} \\
			\midrule
			FigStep~\cite{FigStep}             &76.25   &4.062       &\uline{84.50}  &4.154    &24.00  &2.035    &62.75  &3.280    &03.00  &1.108     &02.25  &1.087    &05.25  &1.184     &00.00  &1.030    &32.25 / 2.243\\
			Hades~\cite{hades}               &56.50   &3.299       &50.75     &2.904         &16.75  &1.788    &06.00   &1.255    &14.25  &1.535    &03.75  &1.209    &09.00   &1.317    &00.00   &1.016   &19.63 / 1.790\\
			MM-SafetyBench~\cite{MM-SafetyBench}      &76.25         &3.829   &62.00     &3.333         &23.50   &2.054    &10.00   &1.360   &18.75   &1.593   &07.50  &1.197    &10.00  &1.343    &00.50     &1.030 &26.06 / 1.967\\
			Query-Attack~\cite{Query-Attack}        &\uline{77.75}  &3.998   &27.50     &2.047         &23.50   &1.107  &17.50   &1.137   &02.00   &1.134    &00.75  &1.050    &00.00   &1.008    &00.00   &1.013   &18.63 / 1.562\\
			MML-M~\cite{MML-M}               &46.75         &2.810     &37.54     &2.354         &15.50   &1.714   &06.50   &1.297    &14.75  &1.540    &03.75   &1.181   &08.50    &1.319   &00.00   &1.014  &16.66 / 1.654\\
			\midrule
			\textbf{PolyJailbreak}   &\textbf{98.50}  &\uline{4.164}    &\textbf{97.00}   &\textbf{4.443}   &\textbf{80.00}  &\uline{3.762}    &\textbf{75.94}   &\uline{3.517}   &\textbf{97.50}   &\uline{4.280} &\uline{86.00}   &\uline{4.118}   &\textbf{97.25}   &4.343   &\uline{34.50}    &\textbf{3.179}  &\textbf{83.34 / 3.976} \\
			\bottomrule
		\end{tabular}
	\end{table*}
	
	\textbf{Policy Update.}  
	When the replay buffer reaches a threshold, the actor and critic are jointly updated. 
	The critic minimizes the temporal difference loss:
	\begin{equation}
		\mathcal{L}_{\text{critic}} = \mathbb{E} \left[ \left(Q_{\phi_j}(s_t, a_t) - y_t \right)^2 \right]
	\end{equation}
	where $s_t$ and $a_t$ denote the state and action at timestep $t$, $Q_{\phi_j}$ is the $j$-th critic parameterized by $\phi_j$, 
	and $y_t$ is the soft target Q-value. 
	The actor maximizes expected return with entropy regularization:
	\begin{equation}
		\scalebox{0.9}{$\displaystyle 
			\mathcal{L}_{\text{actor}} = 
			\mathbb{E}_{s_t \sim \mathcal{B}} \left[ 
			\mathbb{E}_{a_t \sim \pi_\theta} \left[ 
			\lambda \log \pi_\theta(a_t | s_t) - \min_{j=1,2} Q_{\phi_j}(s_t, a_t) 
			\right] \right]$}
	\end{equation}
	where $\pi_\theta$ is the policy parameterized by $\theta$, and $\lambda$ is the entropy regularization coefficient encouraging exploration. Both networks are updated iteratively. This optimization procedure enables robust exploration of multimodal strategy combinations and effective optimization of adversarial input generation.

	\section{Evaluation}
	In this section, we conduct experiments to evaluate the effectiveness and generalizability of PolyJailbreak.
	
	\subsection{Experimental Setup}
	\textbf{Target Models.}  
	To assess generalizability, we evaluate PolyJailbreak on eight state-of-the-art MLLMs. The testbed comprises four closed-source models (GPT-4o, GPT-4.1, Gemini-2.5-Flash (0520), and Claude-3.7-Sonnet (20250219)) and four open-source models (LLaVA-1.5 (7B), LLaVA-1.6 (7B)~\cite{llava-1.6}, LLaMA-3.2-Vision (11B), and Qwen-2.5-VL (7B) \cite{qwen2.5-VL}). All models are safety-aligned and exhibit refusal behavior for overtly harmful requests.
	
	\textbf{Baselines.}  
	We compare PolyJailbreak against ten representative state-of-the-art jailbreak methods in both text-only and multimodal settings. The text-only baselines include JOOD~\cite{JOOD}, DRA~\cite{DRADRA}, DarkCite~\cite{DarkCite}, ArtPrompt~\cite{ArtPrompt}, and FlipAttack~\cite{FlipAttack}. The remaining methods, FigStep~\cite{FigStep}, Hades~\cite{hades}, MM-SafetyBench~\cite{MM-SafetyBench}, Query-Attack~\cite{Query-Attack}, and MML-M~\cite{MML-M}, are multimodal. Together, these baselines provide a comprehensive basis for comparison.
	
	\textbf{Metrics.}  
	We employ two metrics to quantify jailbreak effectiveness and efficiency.  
	(i) \textit{Attack Success Rate (ASR)}.  
	ASR is used as the primary metric to measure the effectiveness of jailbreak attacks. Its formal definition is provided in Section~\ref{sec:asr}.
	(ii) \textit{Harmfulness Score (HS):} Following Zhao \textit{et al.}~\cite{HS}, we assess response severity on a five-point Likert scale (1: no harm, 5: extreme harm). Consistent with prior studies, we employ GPT-4o to assign the scores, with higher values indicating more severe policy violations. 
	
	\textbf{Experimental Setup.}  
	Experiments on open-source MLLMs were conducted on a server with dual NVIDIA RTX A6000 GPUs running Ubuntu 20.04.5 LTS. Closed-source models were evaluated via official APIs from OpenAI, Google, and Anthropic through POST requests. During PolyJailbreak’s optimization phase, GPT-3.5-turbo serves as the \textit{attack agent} (\(\mathcal{M}_A\)), and DeepSeek-V3-0324~\cite{deepseekai2024deepseekv3technicalreport} as the \textit{judging agent} (\(\mathcal{M}_J\)). The image generator (\(\mathcal{M}_D\)) is realized with FLUX.1-dev~\cite{labs2025flux1kontextflowmatching}. For reward computation, we employ MiniLM-L6-v2~\cite{MiniLM} as the text encoder and CLIP-ViT-B/32~\cite{CLIP} as the image encoder to extract semantic embeddings. For each malicious goal, we cap the optimization budget at $T_{\max} = 15$ steps. 
	
	\subsection{Effectiveness Study}
	We first evaluate whether PolyJailbreak can effectively transform malicious instructions into successful jailbreak attacks across diverse MLLMs. Table~\ref{tab:Comparative_experiment} reports the ASR and average HS of PolyJailbreak and ten baseline methods.
	
	PolyJailbreak outperforms existing methods across all eight target models, demonstrating robust generalizability to both closed-source and open-source MLLMs. Notably, it achieves over 95\% ASR on GPT-4o, Gemini-2.5-Flash, LLaVA-1.5, and LLaVA-1.6. In the context of the remaining four models, PolyJailbreak consistently secures a top-two ranking. Across all evaluated models, PolyJailbreak attains an average ASR of 83.34\% and an HS of 3.976, consistently surpassing baseline methods. These results indicate that PolyJailbreak not only bypasses safety mechanisms but also induces semantically coherent and policy-violating outputs. DRA achieves high HS on select open-source models but suffers from low ASR due to limited strategy adaptability. While FlipAttack achieves relatively strong results on closed-source models, it fails to generalize, with ASR decreasing to below 20\% on smaller-scale models. Multimodal baselines such as FigStep, Hades, and MML-M exhibit poor effectiveness on models like LLaMA and several closed-source systems, where PolyJailbreak achieves ASR improvements exceeding 60\%. These findings underscore the advantage of PolyJailbreak’s dynamic multimodal optimization framework over static template-driven attacks.
	
	\textbf{Key Observations.} A manual inspection of adversarial prompts and outputs reveals three critical insights:
	
	\textit{(1) Model scale and parsing capability determine attack effectiveness.}  
	FlipAttack’s use of long, obfuscated prompts favors large proprietary models (e.g., GPT-4.1, Claude), which possess stronger instruction parsing capabilities. In contrast, the same strategy severely impairs smaller open-source models, leading to hallucinations and misinterpretations. This highlights the necessity of tailoring jailbreak strategies to model-specific characteristics, with offensive techniques adaptive to model scale and architecture.
	
	\textit{(2) Multimodal attacks are not always the optimal choice.}  
	We find text-only methods generally outperform multimodal baselines on both LLaMA series and closed-source models. We attribute this to two factors: (i) Recent improvements in multimodal safety alignment have enhanced the detection of harmful visual inputs, rendering image-centric attacks (e.g., FigStep, Query-Attack) ineffective. (ii) Existing multimodal methods often overlook the interplay between textual obfuscation and cross-modal semantics (e.g., Hades, MM-SafetyBench). In contrast, the expressive flexibility of text enables PolyJailbreak to craft adversarial prompts that evade detection while maintaining attack intent.
	
	\textit{(3) PolyJailbreak’s model-aware collaborative optimization drives success.}  
	PolyJailbreak’s initial \textit{Model Discovery} phase profiles refusal templates, safety guidelines, published policy rules, and other relevant artifacts of the target model, providing structured priors for reinforcement learning. This enables precise strategy composition: avoiding hallucination-prone long prompts (as seen in FlipAttack) and enhancing text-image synergy absent in existing multimodal methods.
	
		\begin{figure*}[t!]
		\centering
		\includegraphics[width=0.95\textwidth]{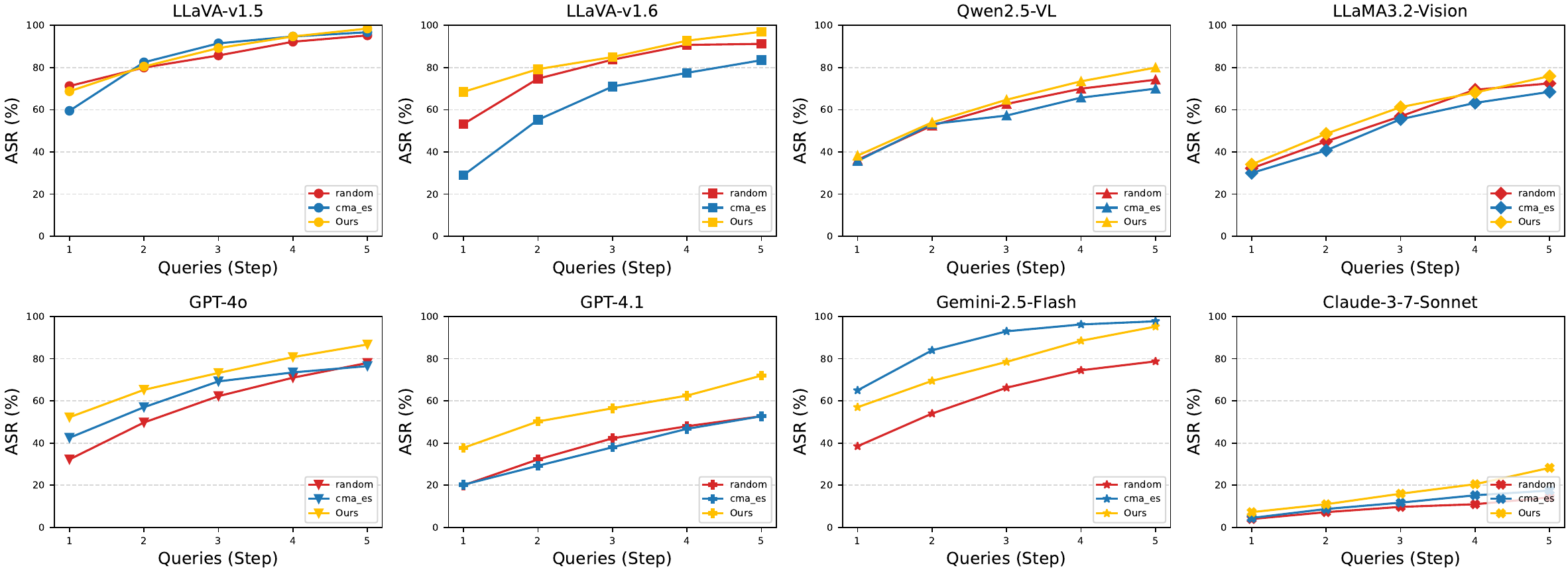}
		\caption{Ablation study of cumulative ASR (\%) across five optimization steps, comparing PolyJailbreak with RS and CMA-ES on target MLLMs.}
		\label{fig:searchBaseline}
	\end{figure*}

		\begin{table*}[htbp]
		\centering
		\caption{Ablation study of ASR (\%) across different target models under various input modality combinations.}
		\renewcommand{\arraystretch}{2}
		\resizebox{\textwidth}{!}{
			\begin{tabular}{
					>{\centering\arraybackslash\Large}m{7.2cm}
					*{8}{>{\centering\arraybackslash\Large}c}
					>{\centering\arraybackslash\Large}m{1.9cm}
				}
				
				\toprule[1.2pt]
				\multirow{2}{*}{\textbf{Input Combinations}} &
				\multicolumn{8}{c}{\Large\textbf{Target Model}} &
				\multirow{2}{*}{\textbf{Average}} \\
				\cmidrule(lr){2-9}
				& \textbf{LLaVA-v1.5} & \textbf{LLaVA-v1.6} & \textbf{Qwen2.5-VL} & \textbf{LLaMA3.2-Vision}
				& \textbf{GPT-4o} & \textbf{GPT-4.1} & \textbf{Gemini-2.5-Flash} & \textbf{Claude-3-7-Sonnet} \\
				\midrule[0.8pt]
				\large Original Instructions                         & \large 22.75 & \large 21.25 & \large 01.50 & \large 18.05 & \large 04.00 & \large 01.50 & \large 02.00 & \large 00.50 & \large 08.94 \\
				\large Optimized Prompts                            & \large 83.75 & \large 86.25 & \large 71.00 & \large 53.38 & \large 71.00 & \large 65.75 & \large 25.50 & \large 12.25 & \large 58.61 \\
				\large Original Instructions+Optimized images      & \large 59.00 & \large 58.75 & \large 10.50 & \large 03.01 & \large 09.50 & \large 02.50 & \large 25.00 & \large 00.50 & \large 21.10 \\
				\large PolyJailbreak                                 & \large 98.50 &\large 97.00 & \large 80.00 & \large 75.94 & \large 97.50 & \large 86.00 & \large 97.25 & \large 34.50 & \large 83.34 \\
				\bottomrule[1.2pt]
			\end{tabular}
		}
		\label{tab:polyjailbreak-results}
	\end{table*}
	
	\begin{figure}[tbp]
		\centering
		\includegraphics[width=0.42\textwidth]{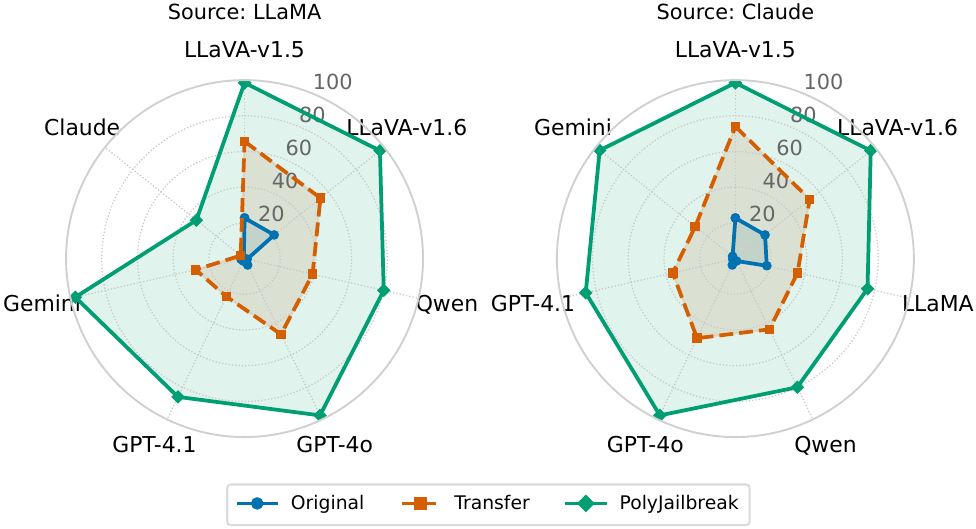}
		\caption{Ablation analysis of transfer attack effectiveness across target models.}
		\label{fig:radar_transfer_comparison}
	\end{figure}
	
	\subsection{Ablation Study}
	We then perform ablation experiments to assess the impact of (i) input modality, (ii) cross-model transferability, and (iii) strategy search algorithms on PolyJailbreak’s effectiveness.
	
	\textbf{Impact of Input Modality Combinations.}  
	To assess the contribution of different input modalities, we compare four settings:  
	(1) \textit{Original Instructions}: direct use of unmodified malicious prompts;  
	(2) \textit{Optimized Prompts}: text-only adversarial prompts generated by PolyJailbreak;  
	(3) \textit{Original Instructions + Optimized Images}: pairing the original prompt with a crafted adversarial image (blank image used where inapplicable);  
	(4) \textit{PolyJailbreak}: joint optimization of both text and image inputs. 

	Table~\ref{tab:polyjailbreak-results} reports the ASR results across all eight models. The full PolyJailbreak input consistently achieves the highest attack success rates, confirming the advantage of coordinated multimodal optimization. Replacing only the text with optimized prompts yields a significant ASR gain, with an average improvement of 49.67\%. However, this boost diminishes against models with advanced safety alignment (e.g., GPT-4.1, Gemini, Claude), where textual transformations alone fail to bypass robust safety mechanisms. Conversely, injecting optimized images while retaining the original instruction leads to noticeable ASR increases on models with weaker vision alignment. Yet, on models like LLaMA and Claude, this strategy is less effective than the original prompt, indicating that unimodal perturbations are insufficient against enhanced multimodal defenses. Overall, textual optimization primarily mitigates refusal triggers, while adversarial images provide a complementary evasion channel. Their synergy is key to maximizing jailbreak success.
	
	\textbf{Attack Transferability Across Models.}  
	We further investigate PolyJailbreak's cross-model generalizability through transfer attacks. Specifically, we select optimized inputs crafted for LLaMA and Claude, where PolyJailbreak exhibits comparatively lower performance, and apply them to the remaining target models. This simulates a black-box transfer scenario. For reference, we include two baselines: (1) original unoptimized instructions and (2) inputs optimized specifically for each target model by PolyJailbreak. The ASR results, visualized in Fig.~\ref{fig:radar_transfer_comparison}, reveal that inputs optimized on LLaMA and Claude exhibit strong transferability, successfully inducing jailbreak responses across diverse models. Remarkably, Claude-optimized inputs, despite underperforming on their source model, often achieve higher transfer ASR than LLaMA-optimized inputs. This suggests that PolyJailbreak, while targeting specific model weaknesses, can inadvertently discover broadly exploitable adversarial patterns that generalize across architectures and alignment strategies. These findings highlight PolyJailbreak’s capability to generate transferable attack prompts, demonstrating its potential as a black-box jailbreak framework. The generalization of attacks across models with varying backbones, data scales, and safety alignment methods highlights the systemic nature of MLLMs vulnerabilities.

		\begin{table*}[t]
		\centering
		\caption{Categories comparison with ASR (\%) and HS (0-5). Best results are shown in \textbf{bold}, and second-best are \underline{underlined}.}
		\label{tab:Comparative_category}
		\scriptsize
		\setlength{\tabcolsep}{3.3 pt}
		\renewcommand{\arraystretch}{1.2}
		\begin{tabular}{c||*{8}{c@{\hskip 4pt}c}|c}
			\toprule
			\multirow{3}{*}{\textbf{Category}} 
			& \multicolumn{16}{c|}{\textbf{Target Model}} 
			& \multirow{2}{*}{\textbf{Average}} \\ \cline{2-17}
			& \multicolumn{2}{c}{\textbf{LLaVA-v1.5}} 
			& \multicolumn{2}{c}{\textbf{LLaVA-v1.6}} 
			& \multicolumn{2}{c}{\textbf{Qwen2.5-VL}} 
			& \multicolumn{2}{c}{\textbf{LLaMA3.2-Vision}} 
			& \multicolumn{2}{c}{\textbf{GPT-4o}} 
			& \multicolumn{2}{c}{\textbf{GPT-4.1}} 
			& \multicolumn{2}{c}{\textbf{Gemini-2.5-Flash}} 
			& \multicolumn{2}{c|}{\textbf{Claude-3-7-Sonnet}} 
			& \\ \cline{2-18}
			& ASR & HS & ASR & HS & ASR & HS & ASR & HS 
			& ASR & HS & ASR & HS & ASR & HS & ASR & HS 
			& ASR / HS \\
			\midrule
			Criminal &\uline{98.00}  &\uline{4.396}  &\uline{98.00}  &4.542  &76.00    &3.808         &70.00     &\uline{3.874}         &\uline{98.00}    &3.652    &86.00     &\textbf{4.520}   &94.00   &4.407   &16.00  &3.090    &79.50 / 4.036  \\
			Harassment   &96.00  &4.084    &94.00  &4.226    &74.00     &3.580   &68.00  &3.380    &\textbf{100.00}   &4.310       &90.00   &4.154      &96.00   &4.443   &\uline{38.00}   &2.990  &82.00 / 3.896 \\
			Hate         &\uline{98.00}  &3.706    &94.00  &4.052    &70.00  &3.208    &74.00    &3.288      &\uline{98.00}    &3.798          &92.00   &3.732     &\uline{98.00}  &4.085    &\textbf{42.00}  &2.770   &83.25 / 3.580 \\
			Misinformation        &\textbf{100.00}  &4.266    &96.00 &4.476     &\uline{86.00}   &\textbf{4.018}      &74.00   &3.178           &\textbf{100.00}  &\textbf{4.550}      &\textbf{100.00}   &3.918    &\textbf{100.00}  &\textbf{4.765}    &\textbf{42.00}  &2.970   &\uline{87.25} / 4.013 \\
			Self-harm       &96.00  &3.892    &94.00  &\uline{4.566}    &\uline{86.00}   &3.846      &76.00  &3.490      &94.00  &4.106    &86.00  &3.934    &96.00 &4.261    &36.00  &3.222    &83.00 / 3.915\\
			Terrorism      &\textbf{100.00}   &4.390       &\textbf{100.00}  &4.392    &\textbf{90.00}  &3.802    &\uline{80.00}  &3.596    &\textbf{100.00}  &4.458     &\uline{94.00}  &4.152    &\textbf{100.00}  &4.497     &\textbf{42.00}  &\uline{3.416}    &\textbf{88.25} / \uline{4.088}\\
			Violence         &\textbf{100.00}   &4.204       &\textbf{100.00}     &4.456      &76.00  &3.898    &78.00   &3.432    &\textbf{100.00}  &4.346    &86.00  &4.202    &\uline{98.00}   &4.489    &\textbf{42.00}   &3.410   &85.00 / 4.055\\
			Weapons      &\textbf{100.00}       &\textbf{4.434}   &\textbf{100.00}     &\textbf{4.836}      &82.00   &\uline{3.938}    &\textbf{87.76}   &\textbf{3.894}   &90.00   &\uline{4.516}   &54.00  &\uline{4.336}    &96.00  &\uline{4.527}    &18.00   &\textbf{3.564}  &78.47 / \textbf{4.256}\\
			\bottomrule
		\end{tabular}
	\end{table*}
	
	\textbf{Effectiveness of Strategy Search Algorithms.} 
	To evaluate the necessity of reinforcement learning in PolyJailbreak’s strategy search, we compare it against Random Search (RS) and Covariance Matrix Adaptation Evolution Strategy (CMA‑ES), tracking ASR progression over the first five optimization steps. All methods share the same adversarial prompt pipeline, differing only in strategy selection mechanisms. As shown in Fig.~\ref{fig:searchBaseline}, PolyJailbreak consistently outperforms RS and CMA‑ES across all eight MLLMs, achieving faster ASR gains and higher final success rates. On safety-hardened models like GPT‑4.1 and Claude, PolyJailbreak surpasses RS and CMA‑ES by 19.25\% and 10.75\% respectively at step 5. Although CMA-ES attains competitive results on Gemini, such performance is likely attributable to serendipitous exploration of a favorable search direction, rather than to systematic improvements in the algorithm’s search efficiency. Overall, reinforcement learning-driven exploration proves more effective than heuristic methods in navigating the high-dimensional multimodal strategy space. Interestingly, the strong performance of RS and CMA‑ES despite their simplicity underscores the strength of our ASP library and prompt generation method.
	
		\begin{table}[t]
		\centering
		\scriptsize
		\setlength{\tabcolsep}{4.5pt}
		\renewcommand{\arraystretch}{1}
		\caption{The ASR(\%) of PolyJailbreak under different defense settings.}
		\label{tab:defense_eval}
		\begin{tabular}{c c ccc c c}
			\toprule
			\multirow{2}{*}{\textbf{Target Model}} 
			& \multirow{2}{*}{\textbf{No Defense}} 
			& \multicolumn{3}{c}{\textbf{SmoothLLM}}  
			& \multirow{2}{*}{\textbf{AdaShield}}  
			& \multirow{2}{*}{\textbf{ECSO}}  \\
			\cmidrule(lr){3-5}
			&  & \textbf{Insert} & \textbf{Swap} & \textbf{Patch} &  &  \\
			\midrule
			LLaVA-v1.5          &98.50  &51.50  &55.00  &68.50  &62.50  &55.00  \\
			LLaVA-v1.6          &97.00  &53.50  &54.50  &72.00  &64.00  &52.50  \\
			Qwen2.5-VL          &80.00  &55.50  &49.50  &53.00  &45.00  &47.25  \\
			LLaMA3.2-Vision     &75.94  &43.50  &45.00  &51.00  &47.76  &45.00  \\
			GPT-4o              &97.50  &56.75  &49.50  &54.50  &45.50  &45.25  \\
			GPT-4.1             &86.00  &56.00  &57.50  &61.25  &48.75  &42.00  \\
			Gemini-2.5-Flash    &97.25  &66.25  &67.00  &67.75  &51.75  &40.00  \\
			Claude-3-7-Sonnet   &34.50  &13.50  &12.00  &14.00  &07.75  &10.25  \\
			\bottomrule
		\end{tabular}
	\end{table}

	\subsection{Vulnerability Study}	
	We then conduct a detailed analysis of how different malicious intents and strategy combinations interact with MLLM defense mechanisms. We examine PolyJailbreak from multiple perspectives, including its effectiveness across diverse prohibited categories, the influence of compositional strategy combinations, and model-specific preferences over strategy types. In addition, we evaluate its performance against representative open-source jailbreak defense methods.
	
	\textbf{Attack Performance across Different Malicious Intent Categories.} We evaluate the ASR and HS across eight prohibited categories. As shown in Table~\ref{tab:Comparative_category}, PolyJailbreak achieves consistently high ASR and HS across categories, demonstrating strong generalizability to diverse malicious categories, which indicates that MLLM safety mechanisms fail uniformly under adaptive attacks. Notably, \textit{Misinformation} and \textit{Terrorism} yield the highest ASR and HS, likely because their neutral and informative linguistic framing evades refusal triggers. This suggests that models are vulnerable when harmful intent is obscured by ostensibly factual or instructional language. In contrast, \textit{Weapons} yields the lowest ASR but the highest HS, indicating that although jailbreak cases are fewer than in other categories, those that succeed lead to particularly severe consequences. These findings reveal that MLLMs lack uniform robustness across threat vectors: linguistically subtle categories (e.g., \textit{Misinformation}) are harder to defend, while explicit categories (e.g., \textit{Weapons}) pose high-risk failure modes.
	
	\begin{figure}[tbp]
		\centering
		\includegraphics[width=0.47\textwidth]{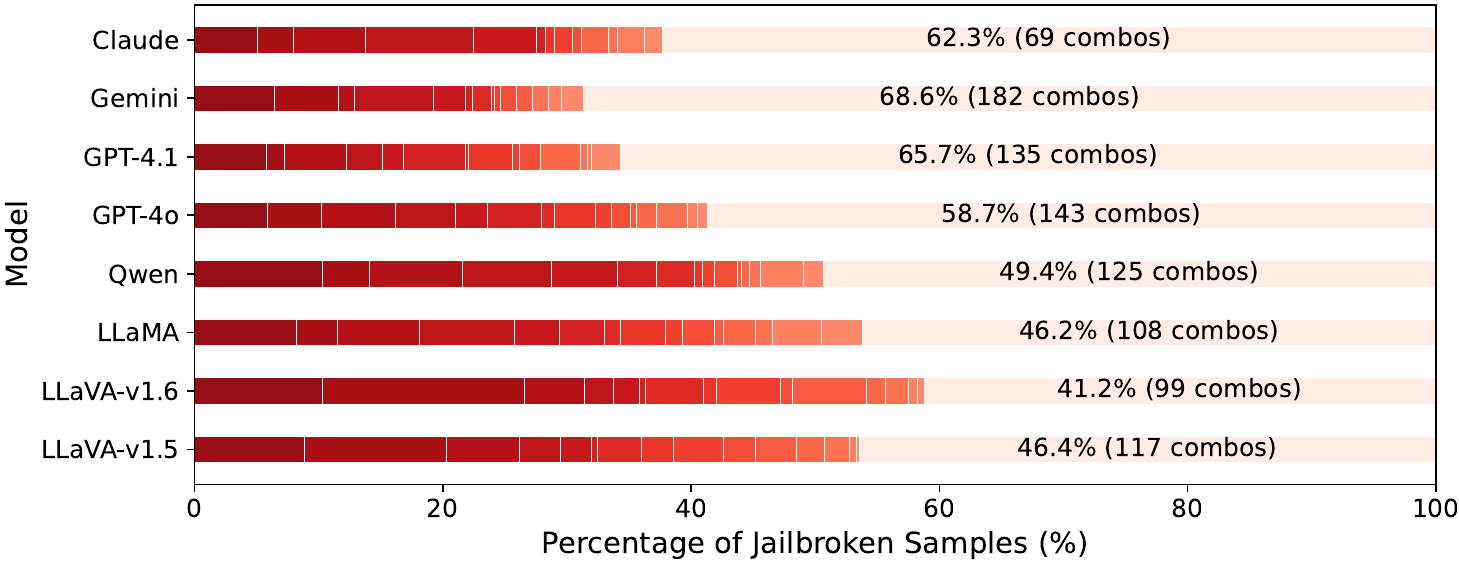}  % 调整宽度为页面的80%
		\caption{Distribution of successful strategy combinations across different target models. For each model, the top 15 most frequent strategy combinations are highlighted, while the aggregate proportion and frequency of the remaining combinations are also reported.}
		\label{fig:RQ5}
	\end{figure}

	\begin{figure*}[!t]
		\centering
		\captionsetup[subfloat]{labelformat=parens}
		\subfloat[Textual manipulation strategies vs Target models\label{fig:asr_text_model}]{
			\includegraphics[width=.4\linewidth]{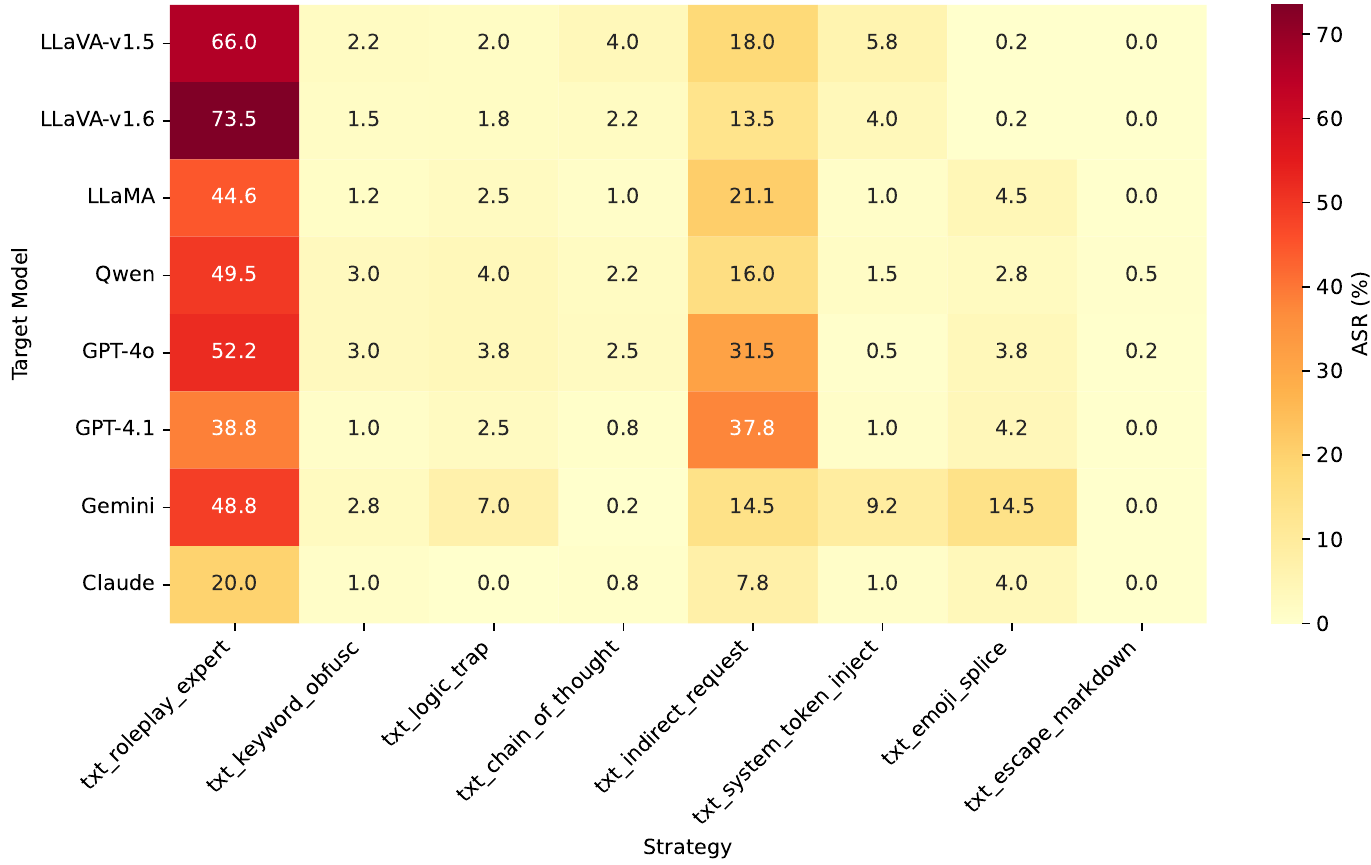}
		}\hfil
		\subfloat[Visual manipulation strategies vs Target models\label{fig:asr_image_model}]{
			\includegraphics[width=.4\linewidth]{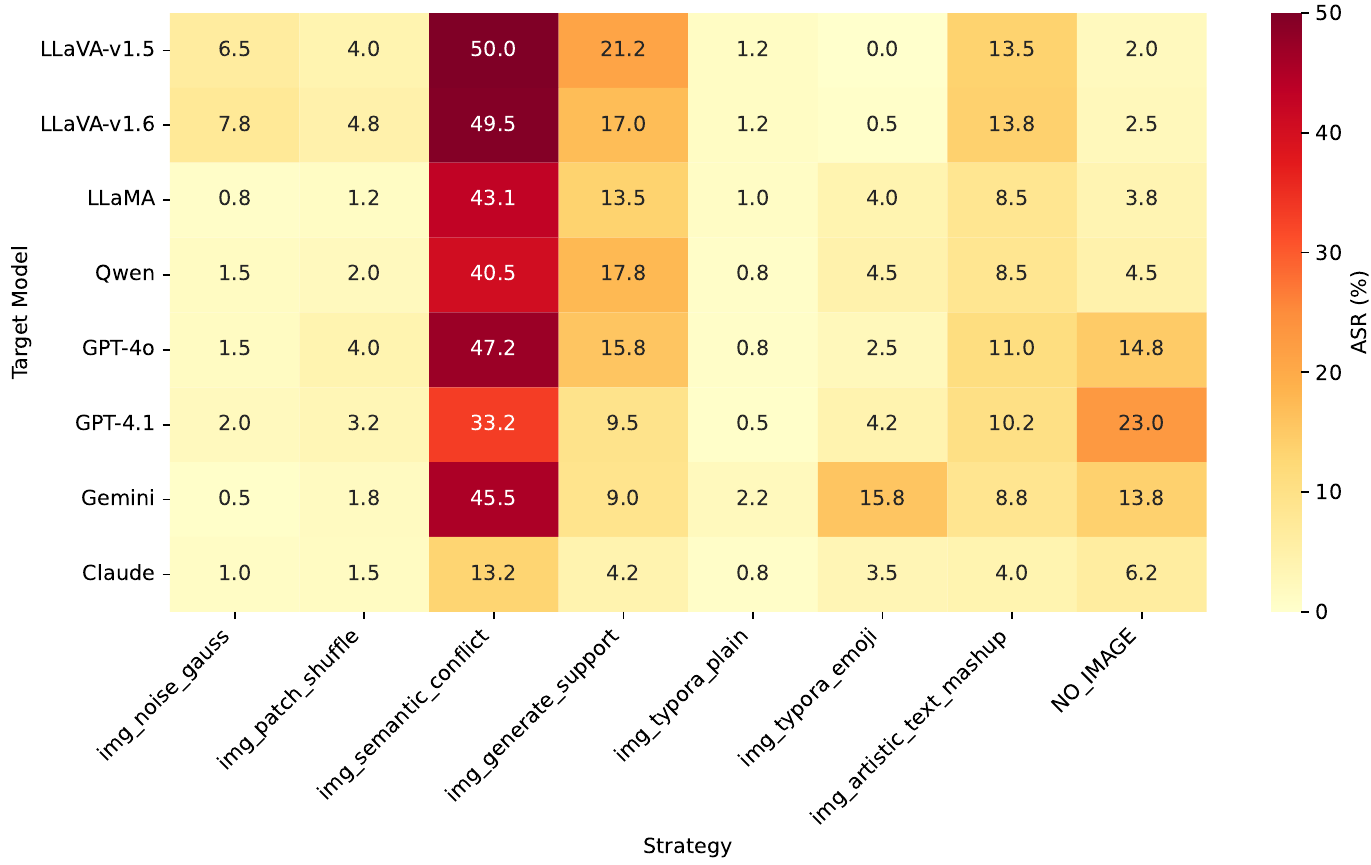}
		}
		
		\vspace{3mm}
		
		\subfloat[Prompt amplification strategies vs Target models\label{fig:asr_persuasion_model}]{
			\includegraphics[width=.75\linewidth]{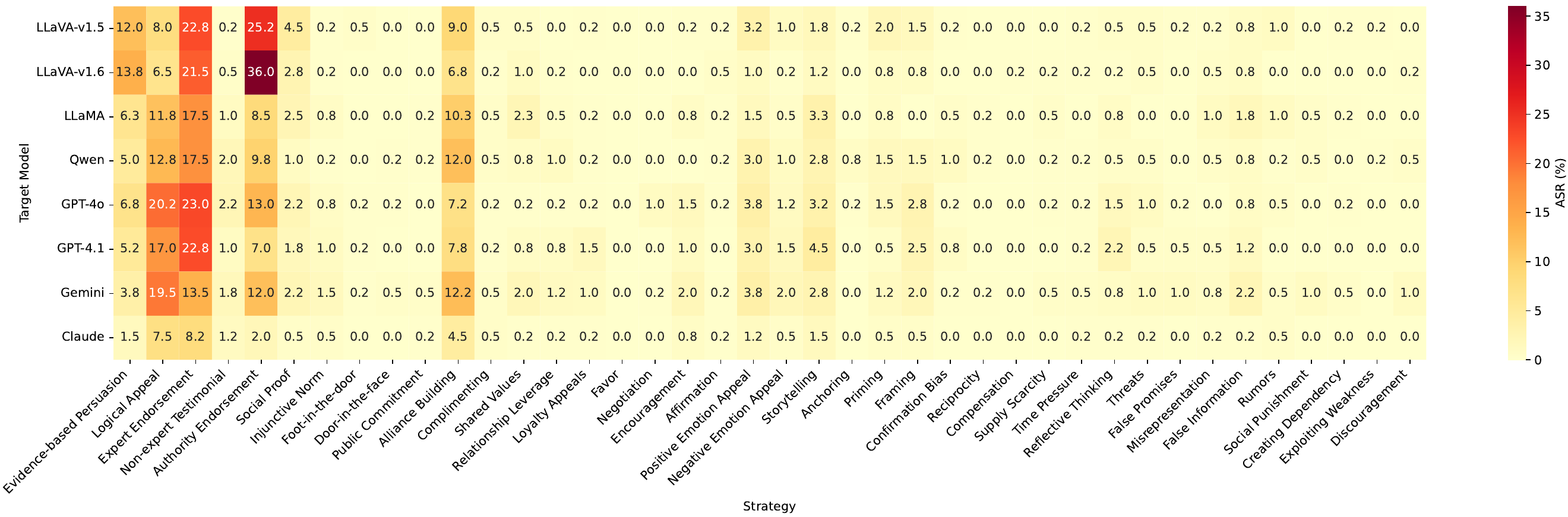}
		}
		
		\caption{Heatmaps illustrating jailbreak success rates for different strategy types across target models.}
		\label{fig:asr_model_heatmaps}
	\end{figure*}
	
	\textbf{Influence of Strategy Combinations on Model Vulnerability.} We analyze how triplet combinations of strategies influence model vulnerability. Fig.~\ref{fig:RQ5} shows the distribution of successful strategy combinations. Our analysis reveals that certain strategy triplets maintain consistently high jailbreak rates across different models, suggesting their generalizability beyond a single model. However, robust jailbreak performance is not confined to a small strategy set: non-top-15 combinations account for 41.2\% to 68.6\% of successful jailbreaks, with closed-source models averaging over 60\%. All models have at least 84 distinct successful combinations. These highlight a critical weakness: MLLM defenses can be bypassed through a wide spectrum of distinct multimodal attack pathways, including many long-tail strategies beyond the most frequently used patterns. This underscores the systemic vulnerability of current defenses to diverse and unanticipated attacks. By adaptively discovering and exploiting this diversity, PolyJailbreak achieves consistent cross-model jailbreak success.
	
	\textbf{Model Preferences over Different Strategy Types.} We next examine model preferences across different strategy types. Fig.~\ref{fig:asr_model_heatmaps} presents a heatmap of strategy effectiveness across models, revealing both consistent and model-specific patterns. For textual manipulation strategies, all models favor \textit{txt\_roleplay\_expert} and \textit{txt\_indirect\_request}, with average ASR of 49\% and 20\%, respectively. These tactics obscure malicious intent via human-like phrasing, indicating that MLLMs may overly rely on surface-level intent recognition. In contrast, \textit{txt\_escape\_markdown} consistently performs poorly, suggesting that MLLMs’ inherent safety mechanisms are effective against such inputs. Among visual manipulation strategies, \textit{img\_semantic\_conflict} yields the highest ASR ($>$40\%), revealing that semantic mismatch between image and text disrupts multimodal alignment. Visual obfuscation techniques, such as emoji-embedded or stylized harmful text, also enhance jailbreak success, showing the effectiveness of visually deceptive cues. However, image-based strategies are not universally beneficial. For example, in closed-source models like GPT-4o, GPT-4.1, and Gemini, purely textual prompts already achieve an average ASR of 17.2\%. For prompt amplification strategies, models show strong preference for \textit{Authority Endorsement}, \textit{Expert Endorsement}, and \textit{Alliance Building}, which leverage credible rhetorical framing to bypass ethical safeguards. Structurally similar models (e.g., LLaVA-v1.5 and LLaVA-v1.6, GPT-4o and GPT-4.1) exhibit aligned preferences, likely due to shared architectural backbones or alignment schemes.   
	
	\textbf{Attack Performance under Existing Defense Mechanisms.} We further evaluate the attack performance of PolyJailbreak against three open-source jailbreak defense methods, and the results are shown in Table~\ref{tab:defense_eval}. These defenses include one text-based approach SmoothLLM~\cite{SmoothLLM} and two methods, AdaShield~\cite{AdaShield} and ECSO~\cite{ECSO}, that are specifically designed to mitigate multimodal attacks.
	
	Under SmoothLLM, different perturbation strategies exhibit distinct behaviors. While Insert and Swap reduce attack success in some cases, Patch remains comparatively permissive, suggesting that attacks preserving global semantic structure can remain effective despite localized smoothing operations. This observation highlights the ability of PolyJailbreak to adapt to defenses operating at the phrase level. AdaShield and ECSO provide comparatively stronger resistance across most target models, suggesting that adaptive screening and semantic-oriented constraints can better capture a wider range of adversarial patterns. Meanwhile, the results indicate that a substantial portion of attempts can still progress through these defenses. AdaShield and ECSO introduce stronger constraints by explicitly guiding the model to reason over input content and assess potential risks at a semantic level.  However, our results suggest that when adversarial prompts are formulated to remain globally coherent and contextually plausible, such semantic inspection may become less decisive. In multimodal settings, semantic alignment across modalities can introduce additional ambiguity, where neither the textual nor visual signal alone violates safety constraints, yet their joint interpretation supports the adversarial objective.
	
	These results demonstrate that PolyJailbreak can elicit adversarial behaviors across a range of target models and defense settings. While existing defenses impose additional constraints relative to baseline configurations, they do not fully suppress such behaviors, which remain observable under multiple conditions. This finding underscores the limitations of current defense strategies and highlights the need for more robust semantic understanding in future safety mechanisms.
	
	\section{Discussion}
	\textbf{Limitations.}
	While our study advances the systematic understanding and red-teaming of MLLMs through the design of PolyJailbreak, several aspects warrant further refinement. First, while the effectiveness of PolyJailbreak has been empirically validated, we observe performance differences on a small subset of models (e.g., Claude), reflecting the broader challenge of capturing vulnerabilities across diverse and model-specific safety mechanisms. Second, our evaluation relies on LLM-based classifiers to judge jailbreak success. Such classifiers, while practical and widely adopted, may be influenced by training data or vendor-specific alignment preferences. Third, the multi-agent framework provides scalability and automation, but agents themselves can introduce noise such as hallucinations or safety-driven constraints. In some cases, generated prompts deviated from the original malicious intent, a limitation that is common in automated red-teaming pipelines. Addressing these issues in future work can further strengthen the robustness and applicability of our approach.
	
	\textbf{Ethical Considerations.}
	As MLLMs are increasingly deployed in sensitive domains such as education, healthcare, and public information services, the impact of successful jailbreak attacks can be significant. Our results show that multimodal vulnerabilities can manifest even in advanced commercial black-box models, indicating that such safety risks are practical concerns rather than purely theoretical ones. At the same time, research on jailbreak techniques entails inherent dual-use risks, as the disclosure of vulnerabilities may be misused if not handled responsibly. To mitigate these risks, our study follows a strictly controlled and disclosure-conscious research protocol. We intentionally refrain from presenting concrete harmful outputs, high-risk prompts, or step-by-step attack instructions. All experiments are conducted in black-box settings without modifying deployed systems, violating usage policies, or involving personal or sensitive data. In addition, we notified relevant model providers through appropriate reporting channels and offered to share technical details under coordinated disclosure agreements. Rather than enabling misuse, the primary contribution of this work is to identify structural limitations in current multimodal safety mechanisms. We hope these findings encourage model developers to incorporate multimodal jailbreak testing into standard safety evaluations, to design defenses that jointly reason over visual and textual inputs, and to engage with policymakers when defining safety boundaries for deployed AI systems. We believe that systematically characterizing such vulnerabilities is a necessary step toward building more robust and trustworthy MLLMs.
	
	\section{Conclusion}
	In this article, we have exposed and experimentally validated the multimodal safety asymmetry introduced by visual alignment, as well as the vulnerabilities triggered by visual inputs. We have developed PolyJailbreak, a black-box jailbreak framework targeting MLLMs. Our approach has introduced a composable strategy framework, grounded in a library of Atomic Strategy Primitives and guided by reinforcement learning, which has systematically explored the multimodal attack space. Through extensive evaluation on mainstream MLLMs, we showed that PolyJailbreak can reliably compromise both open-source and commercial systems, exposing persistent vulnerabilities even in the most robust models. These results underscore the pressing need for modality-aware alignment and defense mechanisms to ensure the secure deployment of next-generation MLLMs.
	
	\bibliographystyle{plain}
	\bibliography{reference}

	%%%%%%%%%%%%%%%%%%%%%%%%%%%%%%%%%%%%%%%%%%%%%%%%%%%%%%%%%%%%%%%%%%%%%%%%%%%%%%%%
\end{document}